\documentclass[proc]{edpsmath}
\usepackage{epsfig,graphicx,setspace,mathrsfs}
\usepackage{amssymb,tikz,wrapfig}
\usepackage[T1]{fontenc} 
\usepackage{txfonts}
\usepackage[mathscr]{euscript}
\graphicspath{{graphics/}}

\newcommand{\Prob}{\mathscr{P}}

\newcommand{\graphset}{{G}}

\newcommand{\N}{{\rm I\!N}}
\newcommand{\R}{{\rm I\!R}}

\newcommand{\vsp}{\vspace*{3mm}}

\newcommand{\bra}{\langle}
\newcommand{\ket}{\rangle}

\newcommand{\order}{{\cal O}}

\newcommand{\one}{\rm 1\!I}

\newcommand{\bc}{\mbox{\boldmath $c$}}

\newcommand{\bxi}{\mbox{\boldmath $\xi$}}

\newcommand{\bz}{\mbox{\boldmath $z$}}

\newcommand{\bJ}{\mbox{\boldmath $J$}}

\newcommand{\bsigma}{\mbox{\boldmath $\sigma$}}
\newcommand{\bomega}{\mbox{\boldmath $\omega$}}

\newcommand{\bphi}{\mbox{\boldmath $\phi$}}

\newcommand{\rme}{\mathrm{e}}
\newcommand{\rmi}{\mathrm{i}}
\newcommand{\rmd}{\mathrm{d}}
\newcommand{\fs}{\footnotesize}

\begin{document}
\title{Random graph ensembles with many short loops}
\author{ES Roberts}
\address{
Institute for Mathematical and Molecular Biomedicine, King's College London,  Hodgkin Building,
London SE1 1UL, United Kingdom} 
\secondaddress{
Randall Division of Cell and Molecular Biophysics, King's College London, New
Hunts House, London SE1 1UL, United Kingdom}

\author{ACC Coolen}
\sameaddress{1} 
\secondaddress{
London Institute for Mathematical Sciences, 35a South St, Mayfair, London W1K 2XF, United Kingdom}

%\email{ekaterina.roberts@kcl.ac.uk, ton.coolen@kcl.ac.uk}

%\pacs{87.18.Vf, 89.70.Cf, 89.75.Fb, 64.60.aq}

%\ead{ekaterina.roberts@kcl.ac.uk ton.coolen@kcl.ac.uk}

\begin{abstract}
Networks observed in the real world often have many short loops. This  violates  the tree-like assumption that underpins the majority of  random graph models and most of the methods used for their analysis.  In this paper we sketch possible research routes to be explored in order to make progress on networks with many short loops, involving old and new random graph models and ideas for novel mathematical methods. We do not present conclusive solutions of problems, but aim to encourage and stimulate new  activity and  in what we believe to be an important but under-exposed area of research. We discuss in more detail the Strauss model, which can be seen as the `harmonic oscillator' of `loopy' random graphs, and a recent exactly solvable  immunological model that involves random graphs with extensively many cliques and short loops. 
\end{abstract}

\begin{resume} 
Les r\'{e}seaux observ\'{e}s dans la Nature ont souvent des cycles courts. Ceci contredit le postulat de hi\'{e}rarchie sur lequel se base la majorit\'{e} des mod\`{e}les de r\'{e}seaux al\'{e}atoires et la plupart des m\'{e}thodes utilis\'{e}es pour leur analyse. Dans cet article, nous esquissons des directions de recherches possibles, afin de progresser sur les r\'{e}seaux contenant beaucoup de cycles courts, faisant appel \`{a} des mod\`{e}les de r\'{e}seaux al\'{e}atoires \'{e}prouv\'{e}s ou nouveaux, et des id\'{e}es pour de nouvelles m\'{e}thodes math\'{e}matiques. Nous ne pr\'{e}sentons pas de solutions d\'{e}finitives, mais notre but est d'encourager et de stimuler de nouveaux travaux dans ce que nous croyons \^{e}tre une direction de recherche importante, bien que insuffisamment explor\'{e}e. Nous discutons en d\'{e}tail le mod\`{e}le de Strauss, qui peut \^{e}tre interpr\'{e}t\'{e} comme `l'oscillateur harmonique' des r\'{e}seaux al\'{e}atoires `à boucles', ainsi qu'un mod\`{e}le immunologique soluble exactement qui implique des r\'{e}seaux al\'{e}atoires avec de nombreux cliques et cycles courts.
\end{resume}
\maketitle

%\tableofcontents

\section{Motivation and background}

Tailored random graph ensembles, whose statistical features are sculpted to mimic those observed in a given application domain, provide a rational framework within which we can understand and quantify topological patterns observed in real life networks. Most analytical approaches for studying such networks, or for studying processes for which they provide the interaction infrastructure, assume explicitly or implicitly that they are locally tree-like. It permits, usually after further mathematical manipulations and in leading orders in the system size, factorisation across nodes and/or links. This in turn allows for the crucial combinatorial sums over all possible graphs with given constraints to be done analytically in the relevant calculations. However, real-world networks - for example protein-protein interaction networks (PPIN), immune networks, synthetic communication networks, or social networks - tend to have a significant number of short loops. It is widely accepted that the abundance of short loops in, for example,  PPINs is intrinsic to the function of these networks. The authors of \cite{Prill} suggested that the stability of a biological network is highly correlated with the relative abundance of motifs (e.g. triangles). The authors of \cite{JeongBerman} and \cite{ElSamad} observed an apparent relationship between short cycles in gene-regulation networks and the system's response to stress and heat-shock.  A highly cited paper \cite{Milo} went as far as to propose that motifs (e.g. triangles) are the basic building blocks of most networks. Similarly, in many-particle physics we know that lattice models are difficult to solve, mainly because of the many short loops that exist between the interacting variables on the lattice vertices \cite{Baxter}; calculating the free energy of  statistical mechanical models on tree-like lattices, in contrast, is relatively straightforward.  It is evident that incorporating constraints relating to the statistics of short loops into the specifications of random graph ensembles is important for this branch of research to be able to more closely align with the needs of practitioners in the bioinformatics and network science communities. 

In this paper we  review  the analytical techniques that are presently available to  model and analyse ensembles of random graphs with extensive numbers of short loops, and we discuss possible future research routes and ideas.  We will use the terms `network' and `graph' without distinction.   We start with a description of tailored random graph ensembles, and argue why graphs with short loops should become the focus of research, to increase their applicability to real-world problems and for mathematical and  methodological reasons. 
We then discuss the simplest network model with short loops that even after some forty years we still cannot solve satisfactorily: the Strauss model \cite{Strauss}, which  is the archetypical ensemble of finitely connected random graphs with controlled number of triangles. We describe some new results on the entropy of this ensemble, continuing the work of e.g. \cite{Burda}, as well as an as yet unexplored approach based on combining graph spectral analysis with the replica method. The next section discusses some recent results on the statistical mechanics of an immune network model \cite{Barra}; in spite of its many short loops, this model could quite unexpectedly be solved analytically with the finite connectivity replica method \cite{VianaBray,finconreplicas,Wemmenhove}, suggesting a possible and welcome new mechanism for analysing more general families of `loopy' graphs. We then show how the model of  \cite{Barra} can indeed be used in other scenarios where short loops in networks play a functional role, and we work out in more detail its application in the context of factor graph representations of protein-protein interaction networks.

\section{Modelling with random graphs -- the imporance of short loops}

Networks are powerful conceptual tools in the modelling of real-world phenomena. In large systems of interacting variables they specify which pairs can interact, which 
leads to convenient visualisations and reduces the complexity of the problem. 
Random graphs  serve as proxies for interaction networks that are (fully or partially) observed or built in biology, physics, economics, or engineering.  Random graphs allow us to analytically solve statistical mechanical models of the processes for which the networks represent the infrastructure, by appropriate averaging of generating functions of observables over all `typical' interaction networks. Or they can be used as `null models' to quantify the statistical relevance of topological measurements which are taken from observed networks. In all cases it is vital that the random graphs actually resemble the true real-world networks in a quantitatively controllable way. 

In this paper we limit ourselves, for simplicity, to nondirected graphs without self-links. Generalisation of the various models and arguments to directed and/or self-interacting graphs is usually straightforward. A nondirected simple $N$-node graph is characterised by its nodes $i\in\{1,\ldots, N\}$ and by the values of $\frac{1}{2}N(N-1)$ link variables $c_{ij}\in\{0,1\}$. Here $c_{ij}=1$ if the nodes $(i,j)$ are connected by a link, and $c_{ij}=0$ otherwise. We always have $c_{ij}=c_{ji}$ (since our graphs are nondirected) and $c_{ii}=0$ (since our graphs are simple), for all $i,j\in\{1,\ldots, N\}$. We will denote the set of all such graphs as $\graphset$. A random graph ensemble is defined by a probability measure $p(\bc)$ on $\graphset$. 

\subsection{Tailored random graph ensembles}

If we wish to use random graph ensembles   to study real-world phenomena, it is vital that our measure $p(\bc)$ favours graphs that mimic those in our application domain. For instance, it makes no sense to use Erd\"{o}s-R\'{e}nyi graphs \cite{ER} as null models against which to test occurrence frequencies of motifs in biological networks: almost any measurement will come out as significant, simply because our yardstick is not realistic.
Tailored random graph ensembles \cite{Alessia,PLOS,RobertsCoolenSchlitt,Handbook} involve measures $p(\bc)$ that are constructed such that specified  topological features of the generated graphs $\bc$ will systematically resemble those of a given real-world graph $\bc^\star\in\graphset$. To construct such measures one first defines the set of $L$ observables $\{\omega_1(\bc),\ldots,\omega_L(\bc)\}$ whose values  the random graphs $\bc$ is supposed to inherit from $\bc^\star$.
One then defines $p(\bc)$ as the maximum entropy ensemble on $\graphset$, subject to the imposition of the values of $\{\omega_1(\bc),\ldots,\omega_L(\bc)\}$, with the Shannon entropy \cite{InfoTheory}  $S[p]=-\sum_{\bc\in\graphset}p(\bc)\log p(\bc)$.  This can be done via hard constraints, where {\em each} $\bc\in\graphset$ with $p(\bc)>0$ must have the specified values, or via soft constraints, where our random graphs will be described by an exponential ensemble and exhibit the specified values of the $L$ observables 
only {\em on average}:
\begin{eqnarray}
{\rm hard~constrained~ensembles:}&~~&p_{h}(\bc)=Z_{h}^{-1}~\prod_{\ell\leq L}\delta_{\omega_\ell(\bc),\omega_\ell(\bc^\star)}
\label{eq:hard}\\
{\rm soft~constrained~ensembles:}&~~&p_{s}(\bc)=Z_{s}^{-1}~\rme^{\sum_{\ell=1}^L\hat{\omega}_\ell \omega_\ell(\bc)},~~~~~
\sum_{\bc\in\graphset}p_{s}(\bc)\omega_\ell(\bc)=\omega_\ell(\bc^\star)~~~\forall\ell\leq L
\label{eq:soft}
\end{eqnarray}
The maximum entropy formulation is essential to make sure one does not introduce any unwanted bias into our tailored graphs; we want to build in the features of $\bc^\star$ and nothing else. 
\hspace*{53mm}
\begin{figure}[t]
\unitlength=0.16mm
\begin{picture}(500,280)
\put(245,130){ {\oval(500,270)}}
\put(290,235){\footnotesize\em all nondirected graphs}
\put(210,100){ {\oval(420,200)}}
\put(240,170){\footnotesize $\omega_1(\bc)\!=\!\omega_1(\bc^\star\!)$}
\put(175,75){ {\oval(340,140)}}
\put(150,115){\footnotesize $\omega_2(\bc)\!=\!\omega_2(\bc^\star\!)$}
\put(95,55){ {\oval(170,90)}}
\put(35,67){\footnotesize $\omega_3(\bc)\!=\!\omega_3(\bc^\star\!)$}
\put(45,30){\fs $\bullet ~\bc^\star$}
\end{picture}
\vsp

\caption{Tailoring of random graph ensembles such that the generated graphs will mimick the features of an observed graph $\bc^\star$, via successive imposition  of the values of $L$ chosen observables $\{\omega_1(\bc),\ldots,\omega_L\}$. Subject to these imposed values, which can be built in as hard constraints (to be reproduced by {\em all} $\bc$ with $p(\bc)>0$), or as soft constraints (to be reproduced on average), the measure $p(\bc)$ is defined by maximising the Shannon entropy $S[p]=-\sum_{\bc\in\graphset}p(\bc)\log p(\bc)$. The smaller the set of graphs that satisfy 
$\omega_\ell(\bc)=\omega_\ell(\bc^\star)$ for all $\ell\leq L$, the more our random graphs are expected to resemble $\bc^\star$.}
\label{fig:boxes}
\end{figure}
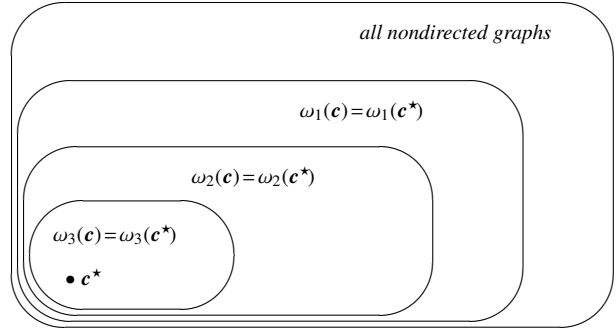

This leads to the question of which would be sensible choices for the observables $\{\omega_1(\bc),\ldots,\omega_L\}$
to carry over from $\bc^\star$ to our ensemble? Sensible choices are those for which we can do the relevant calculations, and for which the Shannon entropy of $p(\bc)$ would be smallest\footnote{Since the effective number of graphs in an ensemble $p(\bc)$ is given by ${\cal N}[p]=\exp(S[p])$, the Shannon entropy can be interpreted as a measure of the size of the smallest box in Figure \ref{fig:boxes}. The smaller this box, the more information on $\bc^\star$ has been carried over to our graph ensemble.}. The  calculations we might wish to do usually relate to stochastic processes for variables placed on the nodes of the graph, and the crucial question is whether the relevant combinatorial sums over all graphs generated from $p(\bc)$ can be carried out analytically. In equilibrium systems we would want to calculate the typical free energy per degree of freedom, averaged over the random graph ensemble, for Hamiltonians of the form 
 $H(\bsigma)=-\sum_{i<j} {c_{ij}}J_{ij}\sigma_i\sigma_j$. The replica method \cite{SK,BOOK,VianaBray,finconreplicas,Wemmenhove} then leads us for hard-constrained ensembles of finitely connected graphs to a combinatorial problem of the following form:
\begin{eqnarray}
\overline{\rme^{-\beta \sum_{\alpha=1}^n H(\bsigma^\alpha)}}&=&
\frac{\sum_{ \bc\in\graphset}\rme^{\sum_{i<j}  {c_{ij}}A_{ij}}\prod_{\ell\leq L}\delta_{\omega_\ell(\bc),\omega_\ell(\bc^\star)}}{\sum_{ \bc\in\graphset}\prod_{\ell\leq L}\delta_{\omega_\ell(\bc),\omega_\ell(\bc^\star)}},~~~~~
A_{ij}=\beta J_{ij}\sum_{\alpha=1}^n  \sigma^\alpha_i\sigma^\alpha_j
\label{eq:statics}
\end{eqnarray}
Similarly, in dynamical studies based on generating functional analysis \cite{Hatchett,Mozeika,Mimura} we would be required to evaluate
\begin{eqnarray}
\overline{\rme^{-\rmi \sum_{it}\hat{h}_i(t)\sum_{j} {c_{ij}}J_{ij}\sigma_j(t)}}&=&
\frac{\sum_{ \bc\in\graphset}\rme^{\sum_{i<j} {c_{ij}}A_{ij}}\prod_{\ell\leq L}\delta_{\omega_\ell(\bc),\omega_\ell(\bc^\star)}}{\sum_{ \bc\in\graphset}\prod_{\ell\leq L}\delta_{\omega_\ell(\bc),\omega_\ell(\bc^\star)}},~~~~~
A_{ij}=-\rmi J_{ij}\sum_t [\hat{h}_{i}(t)\sigma_j(t)\!+\!\hat{h}_{j}(t)\sigma_i(t)]
\label{eq:dynamics}
\end{eqnarray}
In  both cases we see that our observables $\omega_\ell(\bc)$ should be chosen such that sums of the form
$\sum_{ \bc\in\graphset}\delta_{\bomega,\bomega( {\bc})}\rme^{\sum_{i<j}  {c_{ij}}A_{ij}}$ are analytically tractable. Setting $A_{ij}=0$ for all $(i,j)$ gives us  {\em en passant} the value of the Shannon entropy, which for hard-constrained ensembles (\ref{eq:hard}) becomes $S[p]=\log \sum_{\bc\in\graphset}\prod_{\ell\leq L}\delta_{\omega_\ell(\bc),\omega_\ell(\bc^\star)}$. 
It turned out that these summations over all graphs are analytically 
feasible \cite{Alessia,PLOS,RobertsCoolenSchlitt,Handbook}, in leading orders in $N$, for observables such as
\begin{eqnarray}
\bar{k}(\bc)=\frac{1}{N}\sum_{ij}c_{ij},~~~~~~ 
p(k|\bc)=\frac{1}{N}\sum_i\!\delta_{k,\sum_j c_{ij}},~~~~~~
W(k,k^\prime|\bc)=\frac{1}{\bar{k}N}\sum_{ij}\!c_{ij}~\delta_{k,\sum_r c_{ir}}\delta_{k^\prime,\sum_r c_{jr}}
\label{eq:tree_like_observables}
\end{eqnarray}

\subsection{The problem of short loops}

\begin{figure}[t]
 \unitlength=0.55mm
\noindent
 \hspace*{83mm}
\begin{picture}(300,70)
\put(-100,60){$\bc^\star$ = {\em $d$-dim cubic lattice}}
\put(-100,52){$p(k)=\delta_{k,2^d}$} 
 {\put(-95,0){\thicklines{\includegraphics[width=65\unitlength] {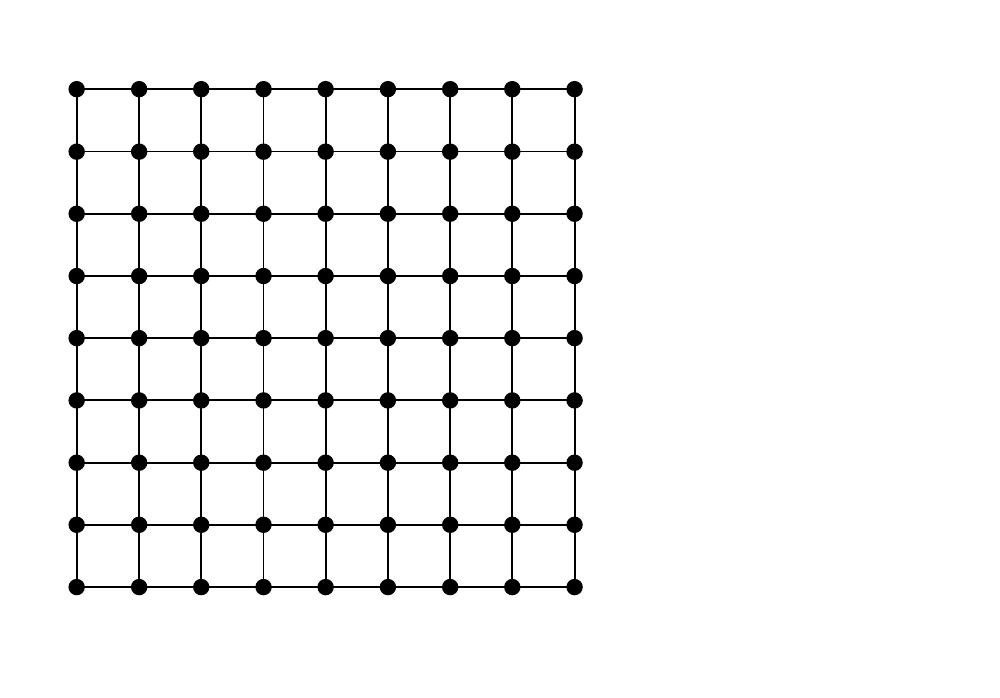}}}}
\put(-7,0){\includegraphics[width=100\unitlength] {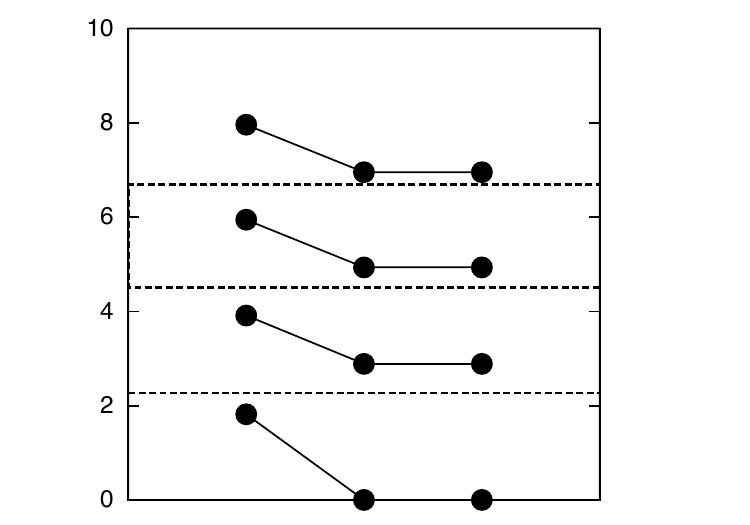}}
\put(25,-5){\fs $A$} 
\put(40,-5){\fs $B$} 
 \put(55,-5){\fs $C$}
\put(-10,55){\small $T_c(d)$}
\put(75,2){\fs $T_c(1)\!=\!0$} 
\put(75,16){\fs $T_c(2)\!=\!2/\!\log(1\!+\!\sqrt{2})$}
\put(75,30){\fs $T_c(3)\!\approx\! 4.512$}
\put(75,44){\fs $T_c(4)\!\approx\! 6.687$}
\end{picture}
\\[10mm]
 \hspace*{83mm}
\begin{picture}(300,65)
\put(-100,60){$\bc^\star$ = {\em`small world' lattice}}
\put(-100,52){$p(k)=\rme^{-q}q^{k-2}\!/(k\!-\!2)!$} 
 {\put(-103,-8){\includegraphics[width=90\unitlength] {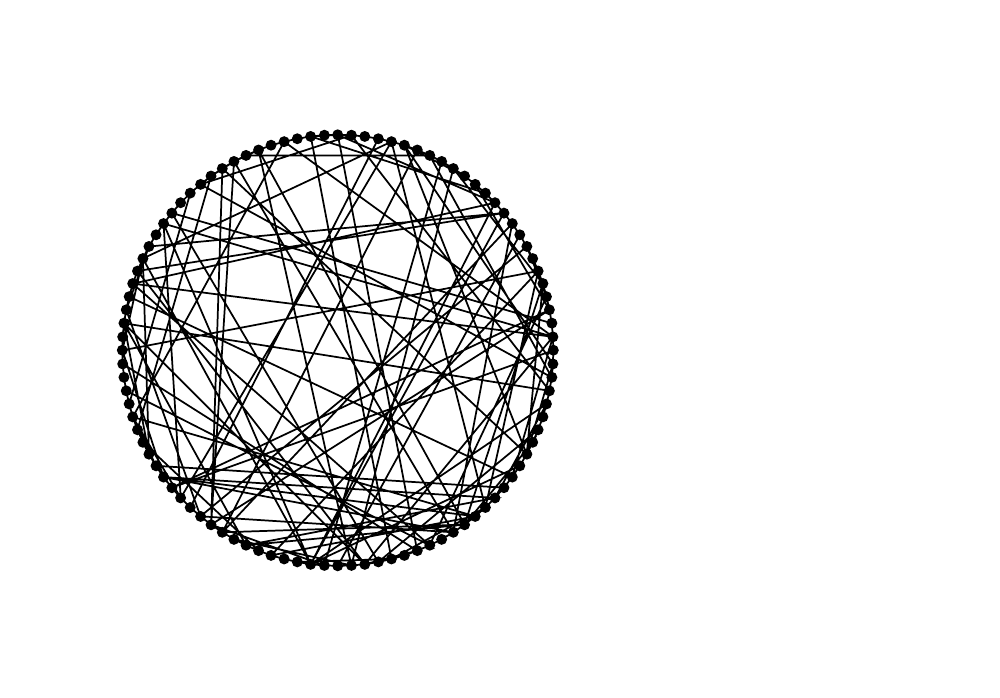}}}
\put(-7,0){\includegraphics[width=100\unitlength] {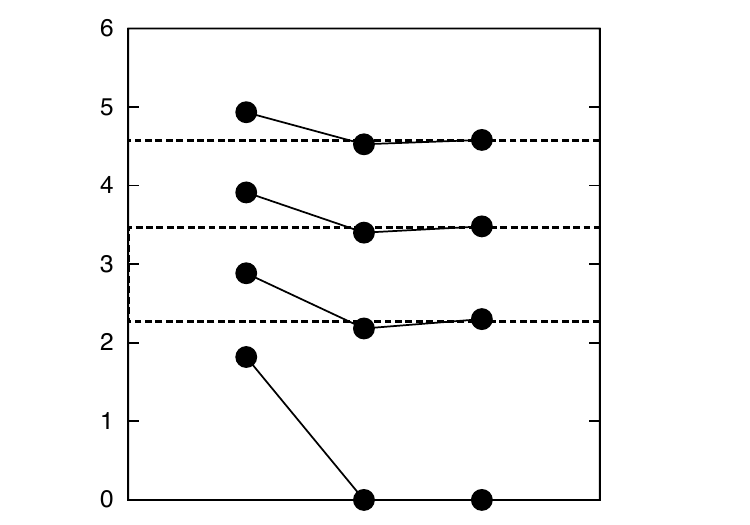}}
\put(25,-5){\fs $A$} 
\put(40,-5){\fs $B$} 
 \put(55,-5){\fs $C$}
\put(-10,55){\small $T_c(q)$}
\put(75,2){\fs $T_c(0)\!=\!0$} 
\put(75,26){\fs $T_c(1)\!=\!2/\!\log(1\!+\!\sqrt{2})$}
\put(75,38){\fs $T_c(2)\!=\!2/\!\log(\frac{3}{4}\!+\!\frac{1}{4}\sqrt{17})$}
\put(75,50){\fs $T_c(3)\!=\!2/\!\log(\frac{2}{3}\!+\!\frac{1}{3}\sqrt{7})$}
\end{picture}
\vspace*{0mm}

\caption{Critical temperatures of Ising systems on finitely connected lattices, calculated for  random graph ensembles (\ref{eq:ensembleA},\ref{eq:ensembleB},\ref{eq:ensembleC}) which are tailored to either resemble $d$-dimensional cubic lattices $\bc^\star$ (top, with $d=1,2,3,4$) or `small world' lattices $\bc^\star$ (bottom, Erd\"{o}s-R\'{e}nyi graph  with 
average degree $q$ superimposed upon a one-dimensional ring, with $q=0,1,2,3$). Connected markers: critical temperatures $T_c(d)$ and $T_c(q)$ calculated analytically for Ising models on the tailored random graphs. Dashed horizontal lines and corresponding values on the right: the true critical temperatures for Ising models on the lattices $\bc^\star$. Transition temperatures are calculated analytically for the small world lattices \cite{SmallWorld} and  for the cubic lattices with $d=1,2$ \cite{Baxter}, and via numerical simulations \cite{IsingSimulations} for cubic lattices with  $d=3,4$.}
\label{fig:lattices}
\end{figure}

To quantify the extent to which real-world networks $\bc^\star$ can be approximated by random graphs that share with $\bc^\star$ the values of the average degree $\bar{k}(\bc^\star)$, or the degree distribution $p(k|\bc^\star)$,  or the joint distribution $W(k,k^\prime|\bc^\star)$ of the degrees of connected node pairs, it is helpful to study systems for which alternative exact solutions or reliable simulation data are available.  Using the methodology of \cite{Alessia,PLOS,RobertsCoolenSchlitt,Handbook}, we can, for instance, calculate with the replica method the critical temperatures of Ising systems on random graphs with Hamiltonian  $H(\bsigma)=-\sum_{i<j}c_{ij}\sigma_i\sigma_j$. We choose our random graph ensembles to be increasingly constrained in the sense of Figure \ref{fig:boxes}:
\begin{eqnarray}
p_A(\bc): && {\rm maximum~entropy~ensemble~with~imposed}~~ \bar{k}(\bc^\star)=\sum_k k~p(k|\bc^\star)
\label{eq:ensembleA}
\\[-1mm]
p_B(\bc):&& {\rm maximum~entropy~ensemble~with~imposed}~~p(k|\bc^\star)~~\forall k\geq 0
\label{eq:ensembleB}
\\[2mm]
p_C(\bc):&& {\rm maximum~entropy~ensemble~with~imposed}~~p(k|\bc^\star)~~{\rm and}~~ W(k,k^\prime|\bc^\star)~~\forall k,k^\prime\geq 0
\label{eq:ensembleC}
\end{eqnarray}
In Figure \ref{fig:lattices} we compare the critical temperatures for Ising systems on random graphs tailored according to 
(\ref{eq:ensembleA},\ref{eq:ensembleB},\ref{eq:ensembleC})  to the true critical temperature values of the approximated finitely connected graphs $\bc^\star$, for cubic lattices and for so-called `small world' lattices. While constraining only the average degree (A) is clearly insufficient, we see that constraining the degree distribution (B) brings us already closer to the true values of the transition temperatures. Adding the joint degree statistics of connected nodes (C) gets the critical temperatures nearly right for the small-world graphs,  but fails to improve $T_c(d)$ for the regular lattices. Since in regular cubic lattices one simply has $W(k,k^\prime|\bc^\star)=kk^\prime 2^{-2d}p(k|\bc^\star)p(k^\prime|\bc^\star)$, prescribing the values of $W(k,k^\prime|\bc^\star)$ here indeed gives no information that is not already contained in the degree distribution. 

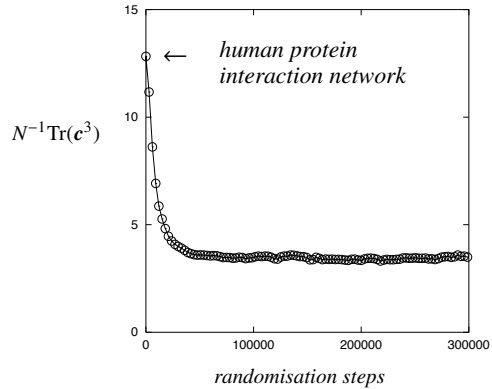
\begin{figure}[t]
\hspace*{0mm}
\unitlength=0.69mm
\hspace*{25mm}\begin{picture}(100,70)
\put(-20,42){\em\fs $N^{-1}{\rm Tr}(\bc^3)$}
\put(0,0){\includegraphics[width=100\unitlength] {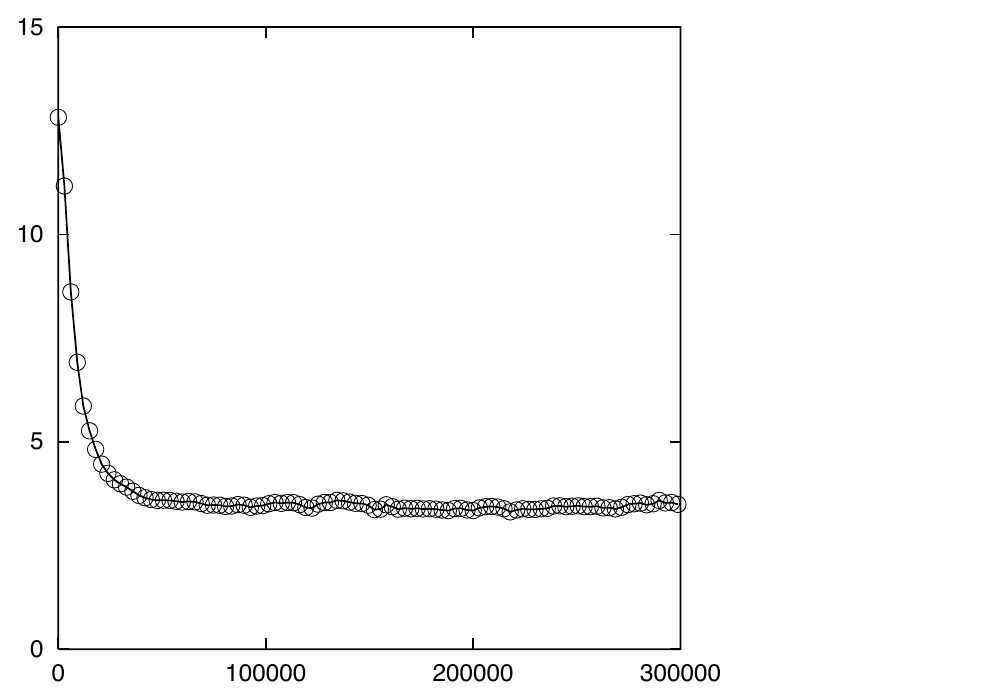}}
\put(19,-5){\fs\em randomisation steps}
\put(20,58){\small\em human protein}
\put(20,53){\small\em interaction network}
\put(9,56.8){$\leftarrow$}
\end{picture}
\vspace*{4mm}

\caption{The effect on the number of triangles per node of randomising the human protein-protein interaction network (PPIN, taken from the HPRD database \cite{hprd}, with $N=9463$ nodes)  within the space of all graphs $\bc\in\graphset$ that have identical degrees and identical joint degree statistics $W(k,k|\bc)$ of connected nodes, generated using the method of \cite{DeMartino}. Note that $N^{-1}{\rm Tr}(\bc^3)=N^{-1}\sum_{ijk}c_{ij}c_{jk}c_{ki}$. Clearly, the human PPIN (the initial state, at steps=0) is atypical in this space, in that it has significantly more triangles per node than expected. }
\label{fig:triangles_in_PPIN}

\end{figure}

We conclude from Figure \ref{fig:lattices} that random graphs tailored on the basis of the observables (\ref{eq:tree_like_observables}) do capture valuable information, and give reasonable approximations of quantitative characterictics of stochastic processes that run on such graphs, but there is room for improvement. A hint at which would be an informative observable to add to $p(k|\bc)$ and $W(k,k^\prime|\bc)$ in order to make our tailored random graphs more realistic approximations of $\bc^\star$ is provided by comparison of the two case studies in Figure  \ref{fig:lattices}. A prominent difference between the topologies of cubic latices and small world graphs is the multiplicity of short loops. Cubic lattices have a finite number per node of loops of any even length, even in the limit $N\to\infty$, whereas in small world lattices the number of short loops per node vanishes for $N\to\infty$. 
Explicit calculation shows that also the ensembles (\ref{eq:ensembleA},\ref{eq:ensembleB},\ref{eq:ensembleC}) typically generate locally tree-like graphs, and this explains why the critical temperatures $T_c(q)$ of the small world graphs are approximated very well, while those of the cubic lattices are not. 
By the same token, one can easily confirm that biological signalling networks, such as protein-protein interaction networks (PPIN) or gene regulation networks (GRN) have signifcantly more short loops than the typical graphs generated within the ensembles 
(\ref{eq:ensembleA},\ref{eq:ensembleB},\ref{eq:ensembleC}). See for example the data in Figure \ref{fig:triangles_in_PPIN}. The same is probably true for many social, economical and technological networks. 

It appears that the next natural observables to be constrained in order to make our tailored random graphs more realistic must involve the number of short loops per node. Moreover, the realistic scaling regime is for this number to be finite, even for $N\to\infty$. However, all available methods for analysing stochastic processes on graphs (replica methods, cavity methods, belief and survey propagation, generating functional analysis),  or for calculating ensemble entropies, all require implicitly or explicitly that the underlying topologies are locally tree-like. Apart from correction methods to handle small deviations from the tree-like assumption \cite{Rizzo,CC,Gomez}, there appears to be as yet no systematic method for doing the relevant combinatorial sums over all graphs $\bc\in\graphset$ in expressions such as (\ref{eq:statics},\ref{eq:dynamics}) analytically when short loops are prevalent.

\section{Strauss model -- the `harmonic oscillator' of loopy graphs}

\subsection{Definitions} 

We now turn to the simplest ensemble of finitely connected graphs with extensively many short loops. Since we have seen earlier that calculating graph ensemble entropies is usually a precursor to  analysing processes on such graphs, we focus for now on how to determine Shannon entropies. 
The Strauss model \cite{Strauss} is the maximum entropy soft-constrained random graph ensemble, with specified values of the average degree and the average number of triangles. 
It is defined via
\begin{eqnarray}
&&
p(\bc) = Z^{-1}(u,g)~ \rme^{u\sum_{ij}c_{ij} + g \sum_{ijk}c_{ij}c_{jk}c_{ki} },~~~~~~~~
Z(u,g) = \sum_{\bc\in\graphset} \rme^{u\sum_{ij}c_{ij} + g \sum_{ijk}c_{ij}c_{jk}c_{ki}} 
\label{eq:Strauss1}
\end{eqnarray}
The ensemble parameters $u$ and $g$ are used to control the relevant ensemble averages, via the identities
\begin{eqnarray}
\langle k \rangle  &=&\sum_{\bc\in\graphset}p(\bc)~\frac{1}{N}\sum_{ij}c_{ij}= \frac{\partial}{\partial u} \phi(u,g)\\
\langle m \rangle  &=& \sum_{\bc\in\graphset}p(\bc)~\frac{1}{N}\sum_{ij}c_{ij}c_{jk}c_{ki}=\frac{\partial}{\partial g} \phi(u,g)
\label{eq:free_energy_to_observables}
\end{eqnarray}
with 
the following function whose evaluation requires that we do analytically the sum over all graphs $\bc\in\graphset$:
\begin{eqnarray}
 \phi(u,g) =& \frac{1}{N} \log Z(u,g) =\frac{1}{N}\log \Big[\sum_{\bc\in\graphset} \rme^{u\sum_{ij}c_{ij} + g \sum_{ijk}c_{ij}c_{jk}c_{ki}} \Big]
\end{eqnarray}
For this ensemble, we wish to calculate the Shannon entropy 
$
S = - \sum_{\bc\in\graphset} p(\bc) \log p(\bc)
$
in leading order in $N$. This follows directly from $\phi(u,g)$, which is minus the free energy of the system, since 
\begin{eqnarray}
\label{eq: S_in_terms_of_phi}
S & = &-  \Big[ \log Z(u,g) - \frac{1}{ Z(u,g) } \sum_{\bc} p(\bc) \log p(\bc) \Big] \nonumber \\
&=& \Big[ 1 - u \frac{\partial}{\partial u} - g \frac{\partial}{\partial g}\Big] \log Z(u,g)
~
 =~  N \left[ \phi(u,g) -u \langle k \rangle  - g \langle m \rangle \right]
\label{eq:strauss_entropy}
\end{eqnarray}
Upon 
setting $g$ to be equal to zero, the ensemble (\ref{eq:Strauss1})  reduces to the Erd\"os-R\'{e}nyi (ER) ensemble \cite{ER}
\begin{eqnarray}
p_{\rm ER}(\bc) = Z_{\rm ER}^{-1}(u) ~\rme^{u \sum_{ij}c_{ij}}, ~~~~~~ 
\label{eq:ER_ensemble}
\log Z_{\rm ER}(u)  =\frac{N(N-1)}{2} \log(\rme^{2u}+1)
\end{eqnarray}  
Differentiating and substituting in  $u=-\frac{1}{2}\ln \left( \frac{1-p}{p}\right)$ then immedately leads us to 
\begin{eqnarray}
\bar{k}=\bra k\ket\big|_{g=0}=
\frac{1}{N}\frac{\rmd}{\rmd u}\log Z_{\rm ER}(u)= \frac{N-1}{1+\rme^{-2u}} = (N\!-\!1)p
\label{eq:first_appearance_of_p}
\end{eqnarray} 
The parameter $p$ can therefore be interpreted as the likelihood of having a link between any two nodes in the ER ensemble, so for finitely connected graphs $p=\order(N^{-1})$ and $u=-\frac{1}{2}\log N+\order(1)$.  This connection with the ER ensemble suggests  rewriting equation (\ref{eq:Strauss1}) as
\begin{eqnarray}
\phi(u,g) &=& \frac{1}{N} \log  \sum_{\bc\in\graphset} p_{\rm ER}(\bc)~ \rme^{ g\sum_{ijk}c_{ij}c_{jk}c_{ki} }
 + \frac{1}{N} \log  Z_{\rm ER}(u) 
\nonumber
\\& =& \frac{1}{2}(N\!-\!1) \log(\rme^{2u}+1)+\frac{1}{N} \log \sum_{r\geq 0}
p(r|u)\rme^{gr}
\label{eq:strauss_free_energy2} 
\end{eqnarray}
with 
\begin{eqnarray}
p(r|u)&=& \sum_{\bc\in\graphset} p_{\rm ER}(\bc)~\delta_{r,\sum_{ijk}c_{ij}c_{jk}c_{ki} }
\label{eq:triangle_stats}
\end{eqnarray}
Hence the substance of the entropy calculation problem for the Strauss ensemble is mathematically equivalent to determining the moments of the distribution $p(r|u)$  of triangle counts in the Erd\"os-R\'{e}nyi ensemble.

\subsection{Simple approximation of the Shannon entropy of the Strauss model}

\begin{figure}[t]
\unitlength=0.77mm
\hspace*{10mm}\begin{picture}(100,60)
\put(0,0){\includegraphics[width=100\unitlength] {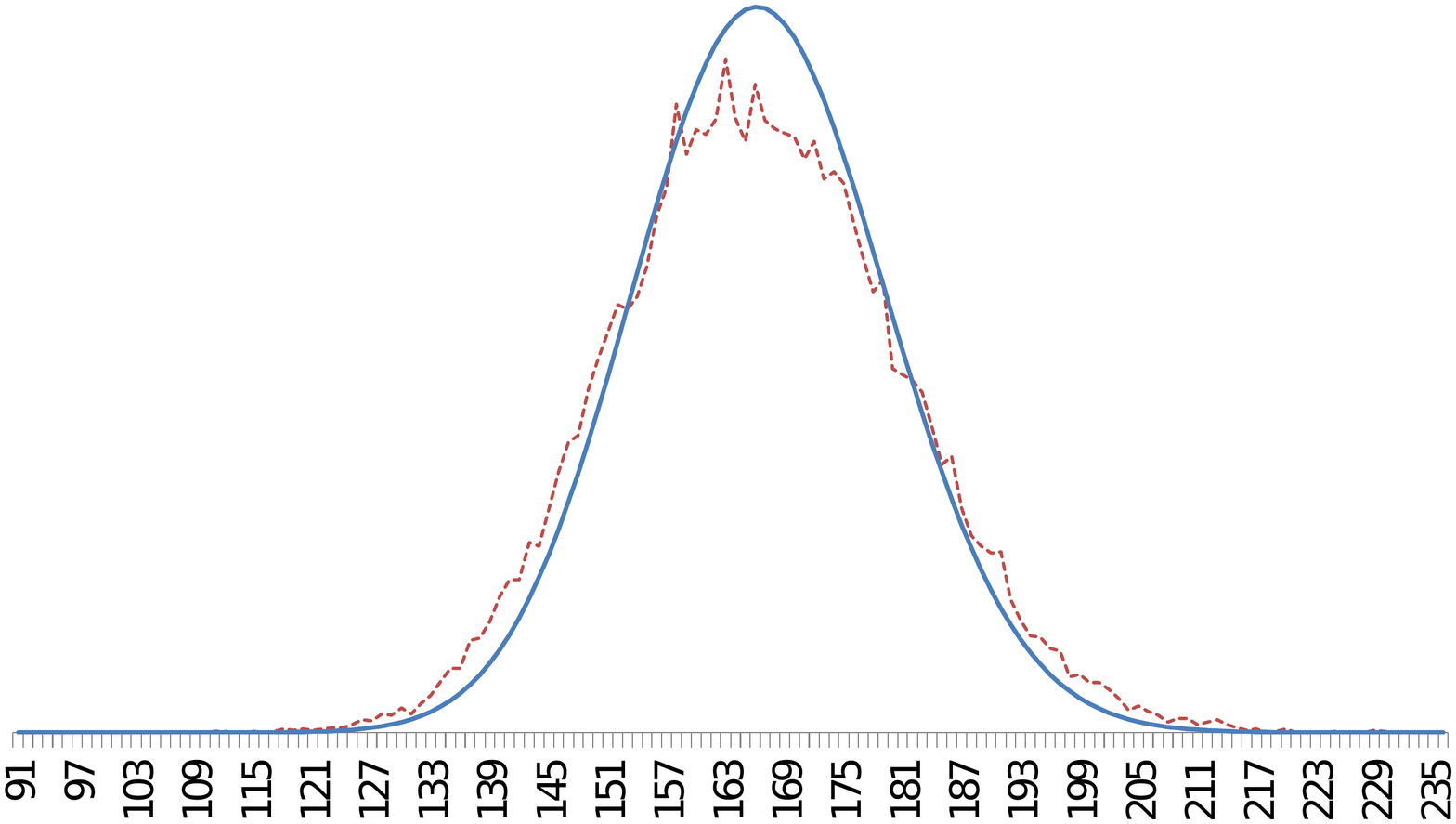}}
\put(-10,50){$p(r)=\big\bra \delta_{r,\sum_{ijk}c_{ij}c_{jk}c_{ki}}\big\ket$}
\put(50,3){$r/6$}
\end{picture}

\caption{Red dotted line: distribution of triangle counts based on sampling 20,000 networks from an Erd\"os-R\'{e}nyi (ER) ensemble with $N=1000$ nodes and $\bar{k}=10$. Dark blue solid line: a Poissonian distribution $p(r)=\rme^{-\bar{r}}\bar{r}^r/r!$ with average number of triangles $\bar{r}$ identical to that measured in the simulated ER graphs. Although similar in shape, the observed triangle distribution $p(r)$ appears to decay to zero more slowly than the Poissonnian one, as $r$ moves away from its average value. }
\label{fig:triangle_distribution_Erdos_renyi}
\end{figure}

We note that the average $\bar{r}(u)$ of the distribution (\ref{eq:triangle_stats})  can be calculated easily and expressed in terms of the average degree $\bar{k}$ of the ER ensemble, giving 
\begin{eqnarray}
\bar{r}(u)&=&\sum_{r\geq 0}r p(r|u)= \sum_{\bc\in\graphset} \sum_{ijk}c_{ij}c_{jk}c_{ki}=N(N\!-\!1)(N\!-\!2)(1\!+\!\rme^{-2u})^{-3}=\bar{k}^3+\order(N^{-1})
\end{eqnarray}
 Here $\bar{k}$, which will differ from the average degree $\bra k\ket$ of (\ref{eq:Strauss1}) as soon as $g>0$, is related to the parameter $u$ via the identity
\begin{eqnarray}
u=-\frac{1}{2}\log[(N\!-\!1)/\bar{k}-1]=
\frac{1}{2}\log(\bar{k}/N)+\order(1/N)
\end{eqnarray}
As a first approximation, we can make the simple ansatz that (\ref{eq:triangle_stats}) is a Poissonian distribution, which must then be given by
$p(r|u)=\rme^{-\bar{r}(u)}[\bar{r}(u)]^r/r!$. One does not expect this assumption to be valid exactly, but according to simulation data (see e.g. Figure \ref{fig:triangle_distribution_Erdos_renyi}), it is a reasonable initial step. It allows us to do the sum over $r\geq 0$ in (\ref{eq:strauss_free_energy2} ), and find
\begin{eqnarray}
\phi(u,g)&=&   
\frac{1}{N}\bar{r}(u)(\rme^g-1)
+\frac{1}{2}(N\!-\!1)\log ( \rme^{2u}+1)
\nonumber
\\
&=&
(N\!-\!1)(N\!-\!2)
(\rme^g-1)(1\!+\!\rme^{-2u})^{-3}
+\frac{1}{2}(N\!-\!1)\log ( \rme^{2u}+1)
\label{eq:phi_strauss}
\end{eqnarray}
We can now immediately work out (\ref{eq:free_energy_to_observables}) and (\ref{eq:strauss_entropy}):
\begin{eqnarray}
\bra k\ket&=& \frac{N-1}{1+\rme^{-2u}}+6(N\!-\!1)(N\!-\!2)\rme^{-2u}(\rme^g\!-\!1)(1\!+\!\rme^{-2u})^{-4}
\\
\bra m\ket&=& \rme^g(N\!-\!1)(N\!-\!2)(1\!+\!\rme^{-2u})^{-3}
\\
S/N&=& (N\!-\!1)(N\!-\!2)
(\rme^g\!-\!1)(1\!+\!\rme^{-2u})^{-3}
+(N\!-\!1)u
+\frac{1}{2}(N\!-\!1)\log ( 1\!+\!\rme^{-2u})
-u\bra k\ket 
-g \bra m\ket
\end{eqnarray}
At $g=0$ we simply recover the equations of the ER model, with $\bra k\ket=(N\!-\!1)/(1\!+\!\rme^{-2u})$ and 
\begin{eqnarray}
g=0:&&
\bra m\ket= \bra k\ket^{3}/N+\order(N^{-2}),~~~~~~
S/N=
\frac{1}{2}\bra k\ket [\log (N/\bra k\ket)+1]
+\order(N^{-1})
\end{eqnarray}

For $g>0$ we need to inspect the solutions of our equations with finite positive values  $\bra k\ket$ and $\bra m\ket$, by working out the different possible scalings of the parameter $u$ with $N$. 
In view of what we know about the scaling with $N$ of the correct solution for $u$ at $g=0$, the natural ansatz to consider is
$u\to -\infty$ as $N\to\infty$. 
We now find  that
\begin{eqnarray}
\bra k\ket=\Big[
\rme^{2u} N+6\bra m\ket
-
6(N\rme^{2u})^2\rme^{2u}\Big] [1\!+\!\order(\rme^{2u},N^{-1})],
~~~~~~
 \rme^{-g}=\frac{N^2\rme^{6u} }{\bra m\ket}[1\!+\!\order(\rme^{2u},N^{-1})]
\end{eqnarray}
Solutions with finite $\bra k\ket$ and $\bra m\ket$ for $N\to\infty$ seem to require that $\rme^{2u}N=\order(1)$, giving
\begin{eqnarray}
u=\frac{1}{2}\log [\bra k\ket\!-\!6\bra m\ket]-\frac{1}{2}\log N +\order(N^{-1}),
~~~~~~
g=-3\log[\bra k\ket\!-\!6\bra m\ket]+\log (N\bra m\ket)+\order(N^{-1})
\end{eqnarray}
This solution clearly exists only if $\bra m\ket<\frac{1}{6}\bra k\ket$. The corresponding entropy expression
is found to be the following, which indeed reduces correctly to the ER entropy for $\bra m\ket\to 0$:
\begin{eqnarray}
S/N&=& \frac{1}{2}[\bra k\ket\!-\!6\bra m\ket]\Big(1-\log[\bra k\ket\!-\!6\bra m\ket]\Big)
+[\frac{1}{2}\bra k\ket\!-\!\bra m\ket]\log N +\bra m\ket[1-\log\bra m\ket]+\order(N^{-1})
\end{eqnarray}
At the point where $\bra m\ket\uparrow \frac{1}{6}\bra k\ket$, we see that $u$ would have to become even more negative.  There is no entropy crisis since
\begin{eqnarray}
\lim_{\bra m\ket\uparrow \bra k\ket/6} S/N&=&
\frac{1}{3}\bra k\ket\log N +\frac{1}{6}\bra k\ket[1-\log(\frac{1}{6}\bra k\ket)]+\order(N^{-1})
\end{eqnarray}
Hence there is no evidence for bifurcation to an alternative solution at $\bra m\ket= \frac{1}{6}\bra k\ket$. The failure of this simple route to lead to solutions in the regime $\bra m\ket\geq \frac{1}{6}\bra k\ket$ must therefore be due to the invalidity of the Poissonian assumption for $p(r|u)$.

\subsection{The diagrammatic approach of Burda et al.}

A drawback of the Strauss model \cite{Strauss}, is that it has a condensed phase, where the typical networks have a tendency to form complete cliques, which does not reflect the topology of real networks.  Burda et al \cite{Burda} refined our understanding of this phenomenon, using a diagrammatic approach. They showed that the clustered phase occurred above certain critical values of the parameters, 
and evaluated the free energy and the expectation value of $\bra m\ket$, in their notation, as
\begin{eqnarray}
\log {Z(G, \gamma)} = \gamma (\rme^G -1) + \log Z_{\rm ER},~~~~~~
\bra m\ket = N^{-1}\bar{k}^3 \rme^G 
\label{eq:<T>} 
\end{eqnarray}
where the identification of the different parameter conventions follows from $\gamma =\bar{k}^3/6$ and $G=6g$.  
 We know also from \cite{Burda} that if we make the substitution $ G^\star \log N + \alpha = G$, in which  $G^\star$ and $\alpha$ are functions of $\bar{k}$ but without $N$ dependency, then the perturbation series will break down for a value of $G^\star$ that is strictly less than 1 (the actual breakdown value  was found numerically to be about 0.7). Hence, we can see that within this range $ N^{G^\star -1}\rightarrow 0 ~~~$as$~~~ N \rightarrow \infty$. This means that the number of triangles per node tends to zero in the large $N$ limit, throughout the regime where the perturbation series of \cite{Burda} converges.
One can also show that, in leading order in $N$,  the average degree remains  unchanged in the regime where the perturbation series converges, i.e. $\bra k\ket=\bar{k}+\order(N^{-1})$. 

If we write $p=\bar{k}/N$, we can write the ensemble entropy according to the expansion of \cite{Burda} as
\begin{eqnarray}
S-S_{\rm ER}&=& 
\left[1+ p(1-p) \ln{(\frac{1}{p} -1)} \frac{\partial}{\partial p}- G \frac{\partial}{\partial G} \right] \frac{\bar{k}^3}{6} (\rme^G -1)
\nonumber
\\
&=& \frac{\bar{k}^3}{6}(\rme^G -1) + \frac{1}{2}\ln \left(\frac{1}{p}-1 \right)\bar{k}^3 (1-p)(\rme^G -1) - \frac{G}{6} \bar{k}^3 \rme^G
\end{eqnarray}
Upon expanding the logarithm, and eliminating the parameter $G$ with equation (\ref{eq:<T>}), it then follows that
\begin{eqnarray}
S
&=&
\frac{\bar{k}\left( N-1 \right)}{2} \left[ 1 + \ln\left( \frac{N}{\bar{k}} \right) \right] 
+ 
\frac{\langle T \rangle}{6} \left[ 1 + \ln\left( \frac{N^3}{\langle T \rangle} \right)\right] 
- 
\frac{\bar{k}^3}{6} \left[ 1 + \ln\left( \frac{N^3}{\bar{k}^3} \right) \right] - \frac{3 \bar{k}^2}{4} + \order(\epsilon_N)
\label{eq:Strauss_entropy}
\end{eqnarray}
in which $\bra T\ket=\bra m\ket N$ is the average number of triangles in typicall graphs from the Strauss ensemble, and $\lim_{N\to\infty}\epsilon_N=0$.
This form has similarities with previously derived results, e.g. \cite{Alessia, RobertsCoolenSchlitt}. If $\langle T \rangle = \bar{k}^3$ then the entropy reduces to the Erd\"os-R\'{e}nyi entropy, as expected.  Direct comparisons with expressions obtained via the Poissonnian ansatz are not valid, because the expressions refer to different scalings with $N$  of $\langle m \rangle$. The next step would be to eliminate $\bar{k}$ in favour of the observable $\langle k \rangle$. This is not simple, as it effectively requires the solution of a fourth order equation.

\begin{figure}[t]
\unitlength=0.71mm
\hspace*{-5mm}
\begin{picture}(100,80)
\put(0,0){\includegraphics[width=120\unitlength]{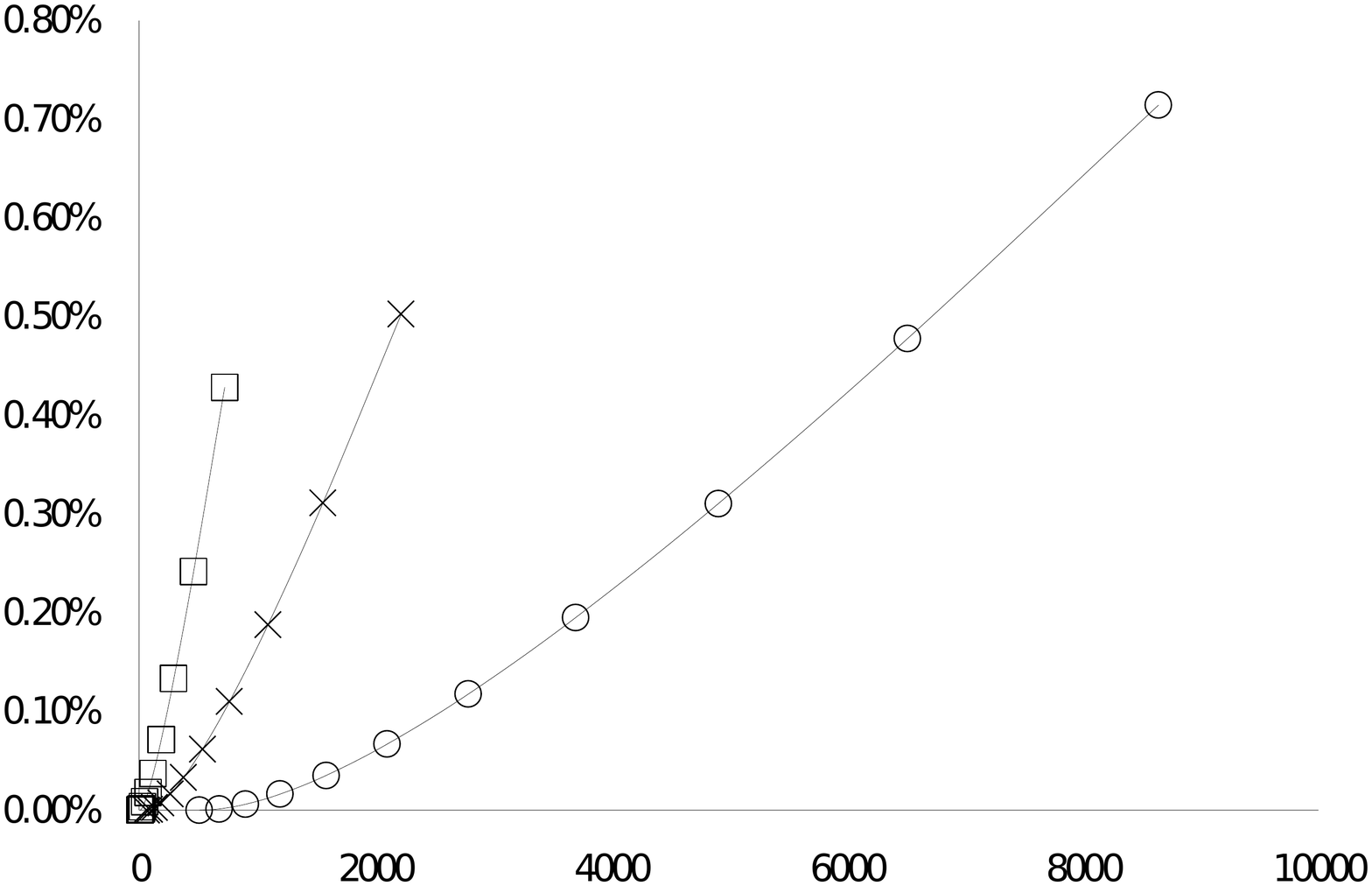}}
\put(-10,50){\large $\frac{S_{\rm ER}-S}{S_{\rm ER}}$}
\put(50,3){\small $\big\bra \sum_{ijk}c_{ij}c_{jk}c_{ki}\big\ket$}
\put(20,51){\small $\bar{k}=2$}
\put(33,56){\small $\bar{k}=4$}
\put(90,72){\small $\bar{k}=8$}
\end{picture}

\caption{Relative complexity (i.e. relative entropy reduction) versus the expected number of triangles in the Strauss model with 
$N=10,000$, calculated from the perturbation theory of \cite{Burda} for values of the coupling constant $G$ below the transition point to the clustered phase. Here $S_{\rm ER}$ is calculated with the actual average degree $\bra k\ket$ of the Strauss model, rather than the implied $\bar{k}$, in order to remove the trivial effect that even a more dense uncorrelated network will automatically have more triangles. }
\label{fig:strauss_complexity}
\end{figure}

\begin{figure}[t]
\unitlength=1.05mm
\hspace*{-5mm}
\begin{picture}(100,78)
\put(0,0){\includegraphics[width=120\unitlength]{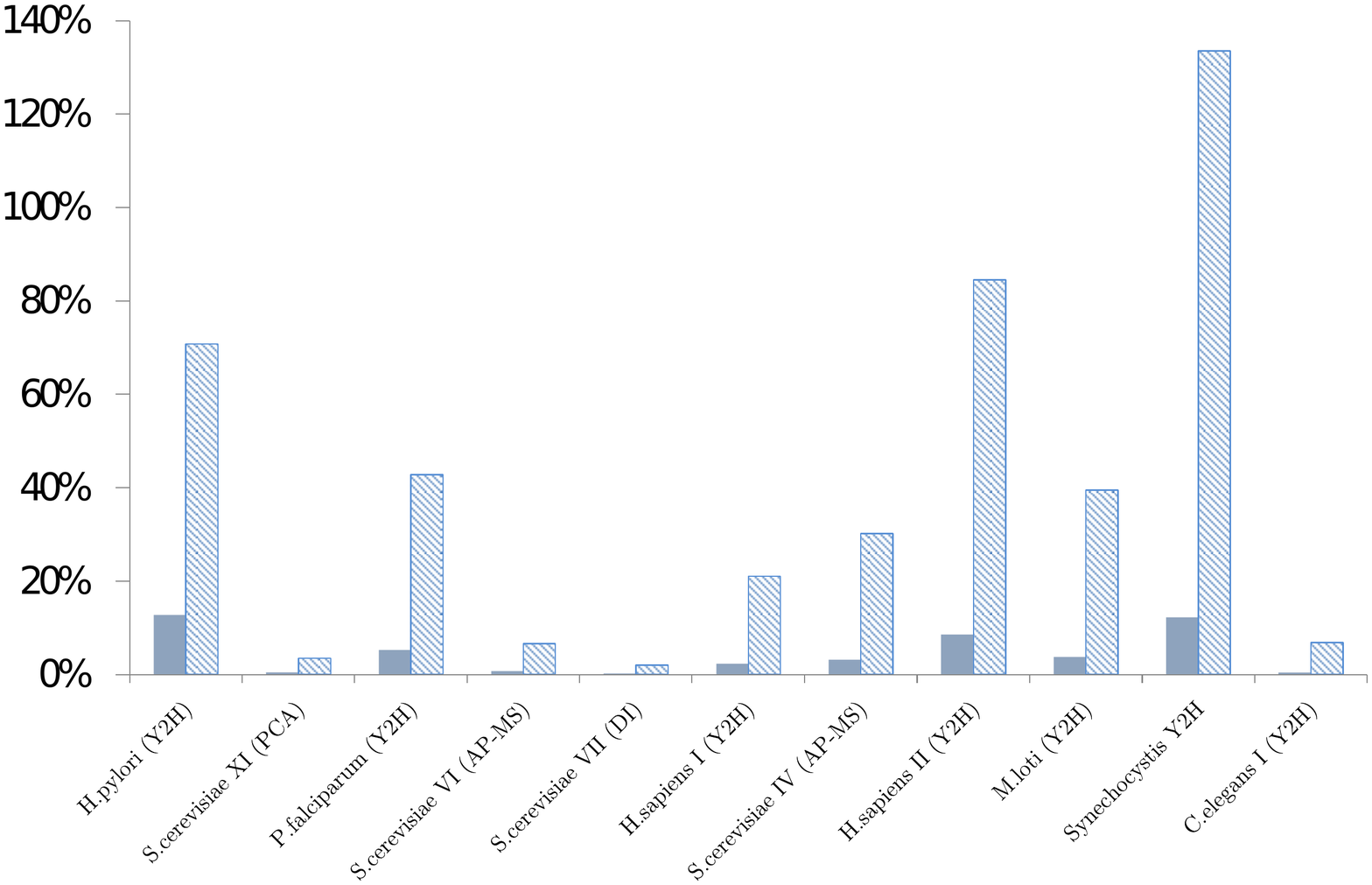}}
\put(-13,60){\large $\frac{\bra m\ket_{\rm ER}}{\bra m\ket_{\rm observed}}$}
\put(-13,50){\large $\frac{\bra m\ket_{\rm Strauss,~max}}{\bra m\ket_{\rm observed}}$}
\end{picture}
\vspace*{-10mm}

\caption{Dark bars: 
number of triangles in an ER ensemble divided by the number of triangles in protein-protein interaction networks (PPIN) of different species, where  both have the same average degree.
Light bars: 
maximum number of triangles in the Strauss ensemble (before the clumping transition) divided by the number of triangles in the observed protein-protein interaction networks, where  both have the same average degree. The species are ordered from left to right in terms of increasing network sizes (which are on average around 2,000 nodes). References for the datasets are  found in\cite{PLOS}. }
\label{fig:strauss_vs_biology}
\end{figure}

The authors of \cite{Burda} numerically deduced the critical values for the coupling parameter $G$ for networks with average degrees of 2, 4 and 8. This gives a region within which we know that it is valid to apply  formula (\ref{eq:Strauss_entropy}), and reasonable to use a model with such parameters to model real networks. We evaluated (\ref{eq:Strauss_entropy}) and related quantities for values of the parameter $G$ up to roughly the critical point.
The implied parameter $\bar{k}$ is found numerically.
 The results are shown in Figure \ref{fig:strauss_complexity}. For the  (realistic) values of average degree and network size considered, we observe that by the time the coupling constant $G$ reaches the critical value, the number of triangles in the network is predicted to increase more than tenfold compared to an ER ensemble with the same average degree. However, the complexity is low when viewed as a proportion of the overall entropy of the ensemble. Constraining only the average number of loops, and remaining within a phase with relatively low clustering,  apparently still leaves significant topological freedom for the networks in the ensemble. 

Figure \ref{fig:strauss_vs_biology} compares the number of triangles observed in several biological networks, with average degree around $\bra k\ket=4$, with the maximum number of triangles that a Strauss ensemble graph could be expected to generate, before it goes into its degenerate clustered phase. This shows that, while the Strauss model is a substantial improvement on the Erd\"os-R\'{e}nyi model from the point of view of the number of loops, it usually collapses into its clustered phase before it reaches biologically realistic values for its parameters. 
If we wish to extend the perturbation analysis beyond the critical point, we need to look at different scalings of the parameter $g$. However, since the un-physical behaviour of the Strauss model above this point has already been shown, such a result would be of limited application. The authors of \cite{Burda} have extended their analysis in \cite{Burda2004} to general uncorrelated degree distributions - but the agreement with simulations was less precise. 

\subsection{Spectrally parametrised loopy ensembles} 

Since the Strauss model `clumps' above a certain critical point before realistic numbers of short loops are achieved, with the imposed triangle numbers being realised in dense cliques,  one would like to define more versatile graph ensembles by including additional observables to tailor the graphs further in terms of short loops and penalise the formation of large cliques. Since the two constrained observables in the Strauss model are both seen to be specific traces of the matrix $\bc$, i.e. 
\begin{eqnarray}
\sum_{ij}c_{ij}=\sum_{ij}c_{ij}c_{ji}={\rm Tr}(\bc^2),~~~~~~\sum_{ijk}c_{ij}c_{jk}c_{ki}={\rm Tr}(\bc^3)
\end{eqnarray}
one in effect contrains in this ensemble the second and third moment of the eigenvalue spectrum 
$\varrho(\mu|\bc)$. 
A natural generalisation of the Strauss ensemble  would therefore be obtained by prescribing more general  spectral features, or even the full eigenvalue  spectrum itself.  For a soft-constrained maximum entropy ensemble this would involve, rather than the scalar pair $(u,g)$, a functional Lagrange parameter $\hat{\varrho}(u)$. Thus we obtain
\begin{eqnarray}
p(\bc) = Z^{-1}[ \hat{\varrho} ] ~\rme^{N \int\! \rmd \mu~ \hat{\varrho}(\mu) \varrho(\mu| \bc)}, ~~~~~~
\varrho(\mu| \bc) = \frac{1}{N} \sum_i \delta\left[ \mu - \mu_i(\bc)\right]
\label{eq:spectrum_formulae}
\end{eqnarray}
where $\mu_i(\bc)$ denotes the $i$-th eigenvalue of $\bc$. For the choice $\hat{\varrho}(\mu) = u \mu^2 + v \mu^3$ one recovers from (\ref{eq:spectrum_formulae}) the Strauss ensemble. Including higher order terms, e.g. via $\hat{\varrho}(\mu)= \sum_{\ell=2}^L v_\ell\mu^\ell$ for some $L\geq 4$, would give better control over the clumping of the original Strauss model. 
If we define our ensemble by a full imposed spectrum $\varrho(\mu)$, which is equivalent to sending $L\to\infty$, we would have to solve the function $\hat{\varrho}(\mu)$ from 
\begin{eqnarray}
\forall \mu\in\R:&& \sum_{\bc\in\graphset}p(\bc) \varrho(\mu| \bc)=\varrho(\mu)
\end{eqnarray}
Although the early steps of the argument in \cite{Burda} would still apply, it is not clear how to analytically re-sum the contributing terms in the expansion for these more general ensembles. Here we need new analytical tools. \vsp

Below we discuss  a potential new route to tackle the relevant combinatorial sums. Determining the entropy of the generalised spectrally constrained graph ensemble (\ref{eq:spectrum_formulae}) would require the evaluation of the functional
\begin{eqnarray}
\phi[\hat{\varrho}]= \frac{1}{N}\log\sum_{\bc\in\graphset}\rme^{N\int\!\rmd \mu~\hat{\varrho}(\mu)\varrho(\mu|\bc)}
\label{eq:spectrum_constraint}
\end{eqnarray}
Calculating the sum over all graphs $\bc\in\graphset$ in  (\ref{eq:spectrum_constraint}) directly is not feasible. However, we can rewrite $\phi[\hat{\varrho}]$ 
using the standard spectrum formula of \cite{Edwards_Jones_1976}:
\begin{eqnarray}
\varrho(\mu|\bc)=\frac{2}{N\pi}\lim_{\varepsilon\downarrow 0}{\rm Im}
\frac{\partial}{\partial\mu} \log Z(\mu\!+\!\rmi\varepsilon|\bc),~~~~~~~~
 Z(\mu|\bc)=
\int_{\R^N}\!\rmd\bphi~\rme^{-\frac{1}{2}\rmi\bphi\cdot
[\bc-\mu\one]\bphi}
\label{eq:Zdef}
\end{eqnarray}
This gives us, after integration by parts in the exponent, 
\begin{eqnarray}
\phi[\hat{\varrho}]&=& \lim_{\varepsilon\downarrow 0}\frac{1}{N}\log\sum_{\bc\in\graphset}\rme^{
-\frac{2}{\pi}{\rm Im} \int\!\rmd\mu ~ \log Z(\mu+\rmi\varepsilon|\bc)~\partial\hat{\varrho}(\mu)/\partial\mu}
\end{eqnarray}
We can now discretize the integral via $\int\!\rmd\mu\to \Delta \sum_\mu$, and use
the identity
$\rme^{-2{\rm Im}\log z}=z^{\rmi}/\overline{z}^{\rmi}$ to write
\begin{eqnarray}
\phi[\hat{\varrho}]&=& \lim_{\varepsilon,\Delta\downarrow 0}\frac{1}{N}\log\sum_{\bc\in\graphset}
\prod_\mu
\rme^{-\frac{2\Delta}{\pi}~(\partial\hat{\varrho}(\mu)/\partial\mu)~{\rm Im} \log Z(\mu+\rmi\varepsilon|\bc)}
\nonumber
\\
&=& 
\lim_{\varepsilon,\Delta\downarrow 0}
\frac{1}{N}\log \sum_{\bc}\prod_\mu \Big[
Z(\mu\!+\!\rmi\varepsilon|\bc)^{\rmi}~
\overline{Z(\mu\!+\!\rmi\varepsilon|\bc)}^{~-\rmi}
\Big]^{\!\frac{\Delta}{\pi}~\partial\hat{\varrho}(\mu)/\partial\mu}
\nonumber
\\
\hspace*{-15mm} 
&=& 
\lim_{\varepsilon,\Delta\downarrow 0}
~
\lim_{n_\mu\to \frac{\Delta\rmi}{\pi}\partial\hat{\varrho}(\mu)/\partial\mu}
~
\lim_{m_\mu\to -n_\mu}
\Phi[\{n_\mu,m_\mu\}]
\label{eq:phi_as_replica_limit}
\end{eqnarray}
in which 
\begin{eqnarray}
\Phi[\{n_\mu,m_\mu\}]&=& \frac{1}{N}\log \sum_{\bc\in\graphset}\prod_\mu\Big[
Z(\mu\!+\!\rmi\varepsilon|\bc)^{n_\mu}~
\overline{Z(\mu\!+\!\rmi\varepsilon|\bc)}^{~m_\mu}
\Big]
\label{eq:bigPhi}
\end{eqnarray}
Expression (\ref{eq:bigPhi}) is clearly reminiscent of formulae encountered in replica analyses of heterogeneous many-variable systems, which suggests a strategy for proceeding with the calculation. We can carry out the sum over all graphs $\bc\in\graphset$, which is the core obstacle in the problem, by evaluating equation (\ref{eq:bigPhi}) for positive integer values of $\{n_\mu,m_\mu\}$ (where the powers of $Z(\mu\!+\!\rmi\varepsilon|\bc)$ and $\overline{Z(\mu\!+\!\rmi\varepsilon|\bc)}$ simply become multiple {\em replicated} Gaussian integrals in which the entries $\{c_{ij}\}$ appear in factorised form). The full expression could then be determined by  taking the limits in (\ref{eq:phi_as_replica_limit}) via analytical continuation. 

The sum over graphs has thereby been tamed, and the previous combinatorial  difficulties converted into the intricacies of an unusual replica limit. In the original papers that launched and used the replica method, the (real-valued) replica dimension $n$ had to be taken to zero, reflecting `frozen' heterogeneity in the micro-parameters of stochastic processes \cite{SK,BOOK,VianaBray,finconreplicas,Wemmenhove}. Later statistical mechanical studies have found how real-valued but nonzero replica dimensions emerge in a natural way to describing nested processes that equilibrate at distinct temperatures and timescales, e.g. slowly evolving heterogeneity in the micro-parameters of `fast' stochastic physical or biological processes  \cite{Penney,Dotsenko,Uezu,Rabello}. To our knowledge there have not yet been calculations in which purely {\em imaginary} replica dimensions emerge, as in the calculation above.

\section{Solvable immune models on loopy networks}

\begin{figure}[t]
\unitlength=0.85mm
\hspace*{-5mm}
\begin{picture}(100,52)
\put(0,0){\includegraphics[width=50\unitlength,height=50\unitlength]{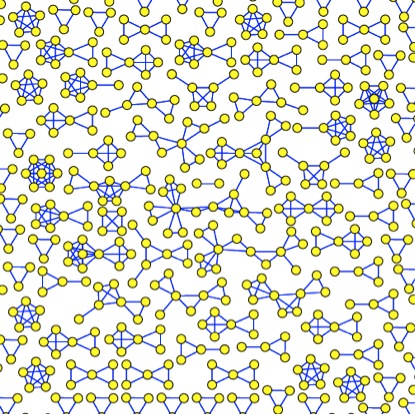}}
\put(60,0){\includegraphics[width=50\unitlength,height=50\unitlength]{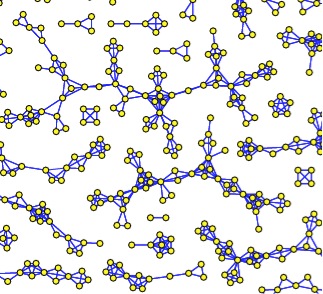}}
\end{picture}
\vspace*{1mm}

\caption{Snapshots of the finitely connected immune network of \cite{Barra}, describing effective interactions between T-clones, that are obtained by integrating out the B-clone variables, for different choices of the model's global control parameters. It is immediately clear from these images  that the graphs of \cite{Barra} have a finite number of short loops per node, and are certainly not locally tree-like. }
\label{fig:immune}
\end{figure}

In a recently published statistical mechanical study of the interaction between T and B-clones in the adaptive immune system \cite{Barra} the authors succeeded in obtaining a full analytical solution,  leading to phase diagrams and testable predictions for observables which agreed perfectly with numerical simulations. Following a preceding paper \cite{Barra_old} they mapped the problem to a new but equivalent spin system describing only T-T interactions, by integrating out the degrees of freedom that represented the B-clones. The new effective model for interacting T-clones could then be solved using replica methods. What is intriguing in the context of this paper is that in this new model for T-clones the spins are positioned on the nodes of a finitely connected interaction graph with an extensive number of short loops, see e.g. Figure \ref{fig:immune}.
Given the arguments in the previous sections, one would not have expected analytical solution to be possible.

\subsection{The model of Agliari et al.}

In the models of \cite{Barra,Barra_old} one studies the interactions between B-clones $b_\mu\in\R$ ($\mu=1\ldots N_B$), T-clones $\sigma_i\in\{-1,1\}$ ($i=1\ldots N_T$), and external triggers of the immune system (the so-called `antigens'). Each B-clone can recognise and attack one specific antigen species, and the T-clones are responsible for coordinating the B-clones via chemical signals (cytokines), but do so in a somewhat promiscuous manner. 
The collective system is described as a statistical mechanical process in equilibrium, characterised by the following Hamiltonian
\begin{eqnarray}
H=\frac{1}{2\sqrt{\beta}}\sum_{\mu=1}^{N_B} {b_\mu^2}-\sum_{\mu=1}^{N_B}{b_\mu}
h_\mu,~~~~~~~~h_\mu=\sum_{i=1}^{N_T}\xi_i^\mu{\sigma_i} +\lambda_\mu{a_\mu}
\end{eqnarray}
Here $a_\mu$ represents the log-concentration of antigen type $\mu$, $\lambda_\mu$ is the sensitivity of the $\mu$-th B-clone to its allocated antigen, and $\xi_i^\mu\in\{-1,0,1\}$ represents the cytokine interaction between T-clone $\sigma_i$ and B-clone $b_\mu$.  The `field' $h_\mu$ acts as the combined input to B-clone $\mu$.  If $h_\mu$ is positive clone $\mu$ will expand, if it is negative clone $\mu$ will contract. The $\xi_i^\mu$ can be excitatory ($\xi_i^\mu=1$), inhibitory ($\xi_i^\mu=-1$), or absent ($\xi_i^\mu=0$), and are drawn randomly and independently from 
\begin{eqnarray}
p(\xi_i^\mu)=\frac{c}{2N_T}\Big[\delta_{\xi_i^\mu,1}+\delta_{\xi_i^\mu,-1}\Big]+(1\!-\!\frac{c}{N_T})\delta_{\xi_i^\mu,0}
\label{eq:cytokines}
\end{eqnarray}
Thus the original system is described by a weighted bi-partite interaction graph. 
The parameter $c\geq 0$ controls the degree of promiscuity of the B-T interactions. 
Realistic clone numbers  would be  $N_B\!\sim\! 10^8$ and  $N_T\sim 10^8$, so statistical mechanical approaches are valid. 
The authors of \cite{Barra} study this system in the regime where  $N_T=N$ and $N_B=\alpha N$, with finite $\alpha>0$ and $N\to\infty$. They 
`integrate out' the B-clones in the system's partition function (which requires only  a simple Gaussian integral),  and are left with a system of interacting Ising spins, described by the effective Hamiltonian
\begin{eqnarray}
H_{\rm eff}({\bsigma})=-\frac{1}{2}\sum_{i,j=1}^{N}{\sigma_i\sigma_j} J_{ij} -\sum_{i=1}^{N}{\sigma_i}\sum_{\mu=1}^{\alpha N}\psi_\mu \xi_i^\mu,~~~~~~~~J_{ij}=\sum_{\mu=1}^{\alpha N}\xi_i^\mu\xi_j^\mu,~~~~~~~~\psi_\mu=\lambda_\mu {a_\mu}
\label{eq:TT}
\end{eqnarray}
This Hamiltonian is reminiscent of the one found in attractor neural network models  \cite{Amit,NNbook}, with a so-called Hebbian interaction marix $J_{ij}$ coupling the spins. However, due  to the scaling with $N$ of the probabilities in (\ref{eq:cytokines}), the present (weighted) interaction matrix is finitely connected. 
Moreover, unlike finitely connected neural network models \cite{Wemmenhove}, where one stores and recalls a finite number of binary patterns with {\em extensive} information content each, here one seeks to store and recall an {\em extensive} number of patterns with a {\em finite} number of bits each. This distinction is not only vital in the immunological context, since an organism has to defend itself against extensive simultaneous invasions to survive, but it also generates fundamental mathematical differences.  Finitely connected attactor models like \cite{Wemmenhove} operate on graphs that are locally tree-like, by construction, whereas the model (\ref{eq:TT}) typically involves `loopy' interaction graphs; see Figure \ref{fig:immune}.

\subsection{Statistical mechanical analysis}

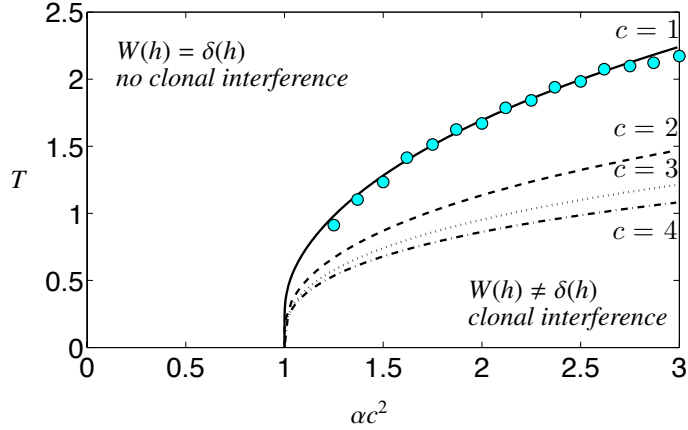
\begin{figure}[t]
\unitlength=0.52mm
\hspace*{5mm}
\begin{picture}(200,98)
\put(2,-5){\includegraphics[width=180\unitlength]{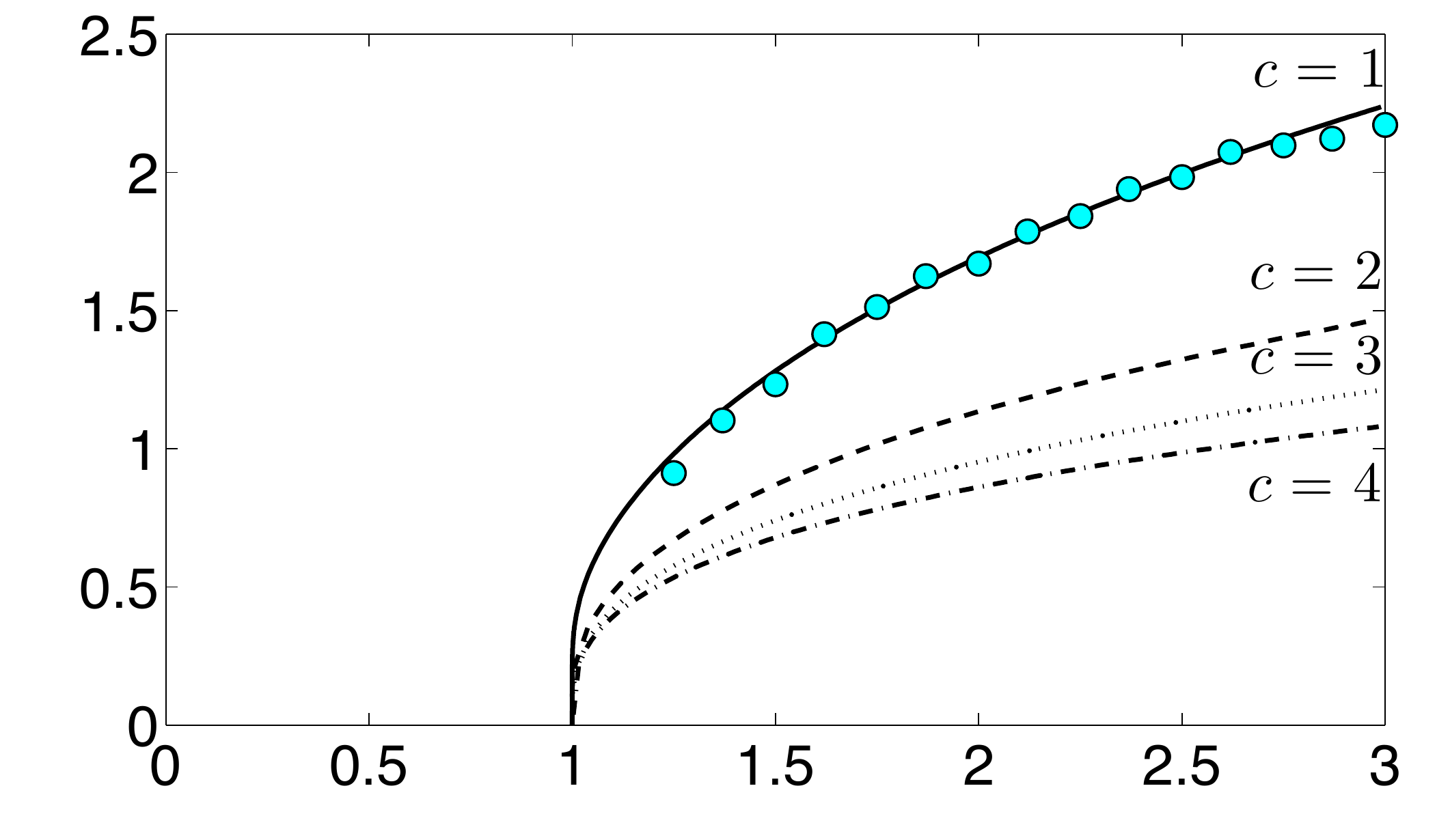}}
\put(90,-10){$\alpha c^2$}
\put(3,50){$T$}
\put(120,22){$W(h)\neq \delta(h)$}
\put(120,15){\em clonal interference}
\put(30,83){$W(h)=\delta(h)$}
\put(30,76){\em no clonal interference}
\end{picture}
\vspace*{8mm}
\caption{Transition lines  in the $(\alpha c^2,T)$ plane for different values of  $c$, with $T=\beta^{-1}$. The distribution $W(h)$ gives the statistics of clonal interference fields, caused by increased connectivity in the graph.  Circles: values calculated via numerical solution of (\ref{eq:RS}) for $c=1$ (diagram reproduced from \cite{Barra}).} 
\label{fig:phase_diagram}
\end{figure}

The details of the statistical mechanical analysis of (\ref{eq:TT}) can be found  in \cite{Barra}, here we only discuss their results. In view of the extensive number of `stored patterns' in this model, compared to finitely connected attractor neural networks, the conventional analysis route (involving sub-lattice magnetisations as order parameters) can no longer be used. Instead the appropriate order parameter is 
\begin{eqnarray}
\Prob(M,\psi)=\frac{1}{\alpha N}\sum_{\mu=1}^{\alpha N} \delta_{M,M_\mu(\bsigma)}\delta(\psi-\psi_\mu)
~~~~~~~~~M_\mu(\bsigma)=\sum_{i=1}^N {\xi_i^\mu} \sigma_i
\end{eqnarray}
Each of the (extensively many) state overlaps $M_\mu(\bsigma)=\sum_{i=1}^N {\xi_i^\mu} \sigma_i$ represent the combined activation/repression signal coming from the T-cells and acting upon B-clone $\mu$, so the conditional distribution $\Prob(M|\psi)$ quantifies the strength and specificity of the response of the adaptive  immune system to a  typical  antigen attack. Given this order parameter, the calculation of the disorder-averaged free energy involves path integral techniques combined with replica methods. The end result, within the replica-symmetric (RS) ansatz, is the following self-consistent equation for a field distribution $W(h)$:
\begin{eqnarray}
W(h)&=&
\rme^{-c}\sum_{k\geq 0}\frac{c^k}{k!}\rme^{-\alpha c k}\sum_{r\geq 0}\frac{(\alpha c)^r}{r!}
\int_{-\infty}^\infty\!\Big[\prod_{s\leq r}\rmd h_sW(h_s)\Big]\sum_{\ell_1\ldots \ell_r\leq k}
\int\!\!\rmd\psi~P(\psi)
\\
&&
\times\!\sum_{\tau=\pm 1}
\delta\left[h\!-\!\tau\psi\!-\!\frac{1}{2\beta}\log\left(\frac{
\sum_{\sigma_1\ldots\sigma_k=\pm 1}
\rme^{\beta(\sum_{\ell\leq k}\!\sigma_{\ell})^2/2c+\beta(\sum_{\ell\leq k}\!\sigma_{\ell})(\psi +\tau/c)+\beta \sum_{s\leq r} h_s \sigma_{\ell_s}}}
{
\sum_{\sigma_1\ldots\sigma_k=\pm 1}
\rme^{\beta(\sum_{\ell\leq k}\!\sigma_{\ell})^2/2c+\beta(\sum_{\ell\leq k}\!\sigma_{\ell})(\psi-\tau/c) +\beta \sum_{s\leq r} h_s \sigma_{\ell_s}}}\right)\right]
\label{eq:RS}
\end{eqnarray}
in which $\beta$ is the inverse temperature (i.e. inverse noise level) of the system. 
 The distribtion $W(h)$ turns out to describe the distribution of clonal interference fields, i.e. the unwanted signalling cross-talk between clones.  
Equation (\ref{eq:RS}) always has the trivial solution  $W(h)=\delta(h)$, which represents interference-free operation, but exhibits bifurcations away from this state at parameter combinations $(\alpha,c,\beta)$ such that
\begin{eqnarray}
1=
\alpha c^2
\sum_{k\geq 0}\rme^{-c}\frac{c^{k}}{k!}
\left\{
\frac{\int\!{\rm d}z~\rme^{-\frac{1}{2}z^2}\tanh(z\sqrt{\beta/c}\!+\!\beta/c)\cosh^{k+1}(z\sqrt{\beta/c}\!+\!\beta/c)
}
{\int\! {\rm d}z~\rme^{-\frac{1}{2}z^2}\cosh^{k+1}(z\sqrt{\beta/c}\!+\!\beta/c)
}
\right\}^2
\label{eq:immune_transition}
\end{eqnarray}
This equation therefore defines a phase transition that separates a regime of interference-free operation of the adaptive immune system  from one that exhibits deteriorating performance due to interference between clones. Cross-sections of this transition surface are shown in Figure \ref{fig:phase_diagram}.

\subsection{Stepping back - why is the `loopy' Agliari et al model solvable?}

In retrospect it is clear why the interacting spin model (\ref{eq:TT})  is solvable, in spite of the extensively many short loops in its interaction graph $\bJ=\{J_{ij}\}$. The reason is that the $N\times N$ interaction matrix has the form $\bJ=\bxi^\dag\bxi$, with a $p\times N$ matrix $\bxi$ with entries $\{\xi_i^\mu\}$ that itself represents a locally tree-like graph. Mathematically this allows one to introduce a Hubbard-Stratonovich-type \cite{Baxter} transformation in the partition function, to a new but equivalent model that has a locally tree-like structure, at the cost of introducing $p$ new Gaussian degrees of freedom:
\begin{eqnarray}
J_{ij}=\sum_{\mu=1}^p \xi_i^\mu\xi_j^\mu:&&
\sum_{\bsigma\in\{-1,1\}^N}\rme^{\beta\sum_{i<j}J_{ij}\sigma_i\sigma_j}
=\int_{\R^p}\!\frac{\rmd \bz}{(2\pi)^{p/2}}\sum_{\bsigma\in\{-1,1\}^N}\rme^{\sqrt{\beta}\sum_{\mu i}z_\mu \xi^{\mu}_ {i}\sigma_i-\frac{1}{2}\sum_\mu z_\mu^2}
\end{eqnarray}
This construction indeed reflects exactly the origin of the model in \cite{Barra}, with the B-cells acting as auxiliary Gaussian variables. The more general question would now be for which other interaction networks with extensively many short loops one could write (or approximate sensibly) the interaction marix in a form $\bJ=\bxi^\dag\bxi$ for some $p$ and some
$p\times N$ bi-partite but locally tree-like graph $\bxi$.  Models on such networks could then be mapped similarly onto an equivalent model where the short loops have been traded in for extra degree of freedom. This would obviously only be possible for graphs with nonnegative spectra. 
\vsp

Another curious aspect of the Agliari et al model is that it defies another expectation. Normally the entropy calculation for a random graph ensemble is easier than solving an interacting spin model on the ensemble's graphs. Here the situation is reversed. The entropy calculation is harder.   
For the present nondirected weighted graph ensemble we have 
\begin{eqnarray}
\bc\in \N^{\frac{1}{2}N(N-1)}=\graphset^\prime,~~~~~~~~p(\bc)=\Big\bra \prod_{i<j}\delta_{c_{ij},\sum_{\mu=1}^{p}\xi_i^\mu\xi_j^\mu}\Big\ket_{\bxi}
\label{eq:separable_ensemble}
\end{eqnarray}
in which $\bra\ldots\ket_{\bxi}$ denotes averaging over (\ref{eq:cytokines}). 
This ensemble is not a maximum entropy ensemble with constraints in the sense of (\ref{eq:hard}) or (\ref{eq:soft}), and the methods used for the latter ensembles are no longer applicable. 
Using the replica identity $\bra \log Z\ket=\lim_{n\to 0}n^{-1}\log\bra Z^n\ket$ we can, however, write the Shannon entropy per node of the ensemble (\ref{eq:separable_ensemble}) as
\begin{eqnarray}
S&=&\sum_{\bc\in\graphset^\prime} p(\bc) \log p(\bc) 
=
\lim_{n \rightarrow 0 } \frac{1}{n}\log  \sum_{\bc\in \graphset^\prime} p^{n+1}(\bc) 
\nonumber
\\
&=&\lim_{n \rightarrow 0 } \frac{1}{n}\log \sum_{\bc\in \graphset^\prime}
\Big\bra\ldots\Big\bra 
\prod_{i<j}\Big[\prod_{\alpha=1}^{n+1}\delta_{c_{ij},\sum_{\mu=1}^{p}\xi_{i,\alpha}^{\mu}\xi_{j,\alpha}^{\mu}}
\Big]
\Big\ket_{\bxi_1}\ldots~\Big\ket_{\bxi_{n+1}}
\nonumber
\\
&=&\lim_{n \rightarrow 0 } \frac{1}{n}\log 
\Big\bra\ldots\Big\bra 
\prod_{i<j}\Big\{\sum_{\ell\geq 0}\Big[\prod_{\alpha=1}^{n+1}\delta_{\ell,\sum_{\mu=1}^{p}\xi_{i,\alpha}^{\mu}\xi_{j,\alpha}^{\mu}}
\Big]\Big\}
\Big\ket_{\bxi_1}\ldots~\Big\ket_{\bxi_{n+1}}
\end{eqnarray}
Working out the leading orders in $N$ of this remaining combinatorial problem looks feasible but has not yet been fully solved, 
and is the subject of ongoing work.

\subsection{The connection with models of proteins and their complexes}
\label{sec:PCN}

\begin{figure}[t]
\unitlength=0.85mm
\hspace*{5mm}
\begin{picture}(100,62)
\put(0,0){\includegraphics[width=80\unitlength]{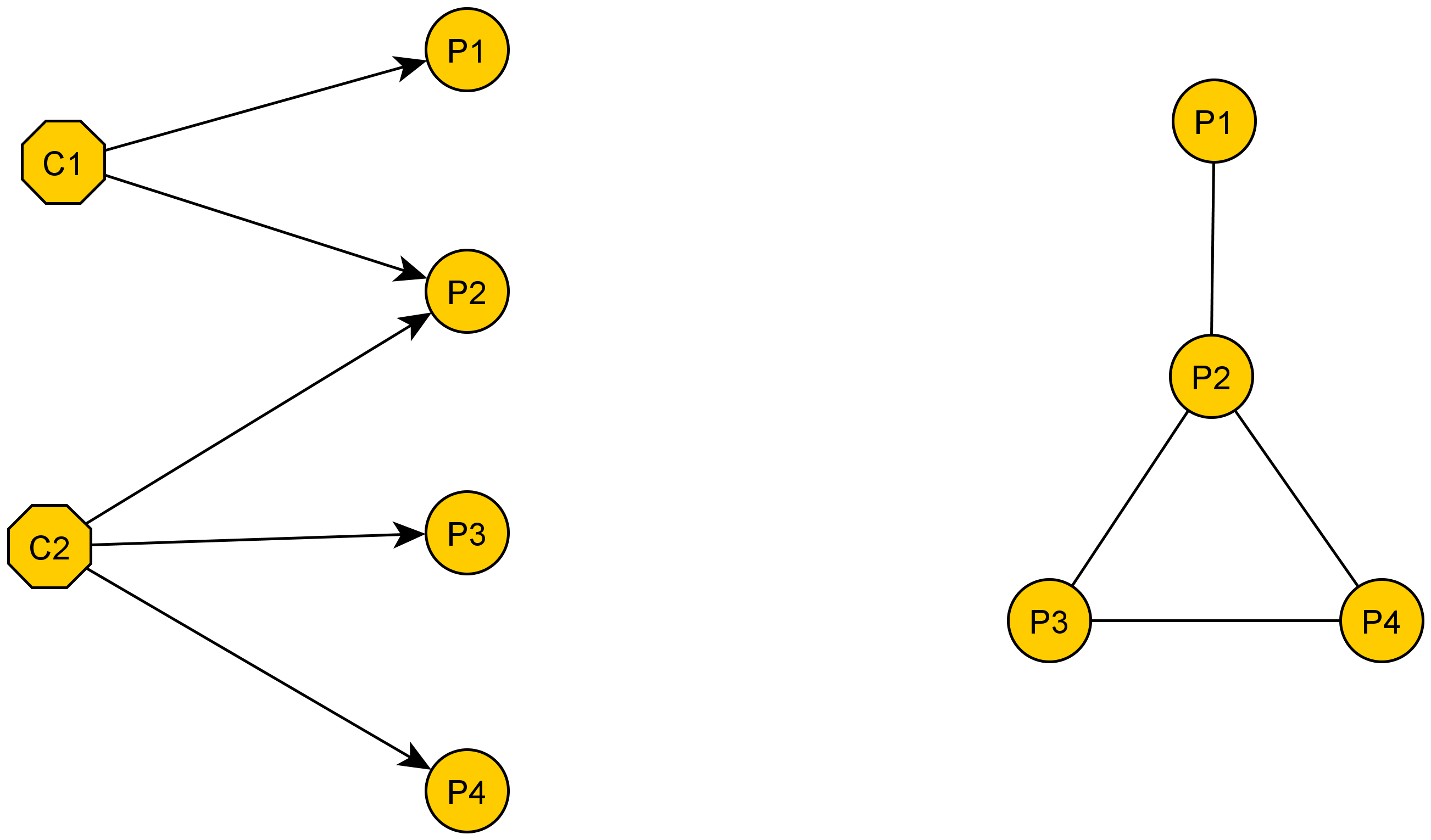}}
\put(-6,54){\em complexes}
\put(20,54){\em proteins}
\put(63,50){\em PPIN}
\end{picture}
\caption{In the bipartite graph on the left proteins and complexes are both represented as nodes, and a link indicates that a given protein is a constituent of the complex.  In the graph on the right (the standard representation of PPIN databases) there is a link between two proteins if they are jointly part of one or more protein complexes.  The latter graph is weighted if the value of the link $c_{ij}$ is defined as the {\em number} of complexes in which proteins $P_i$ and $P_j$ participate simultaneously.}
\label{fig:PPINcomplexes}
\end{figure}
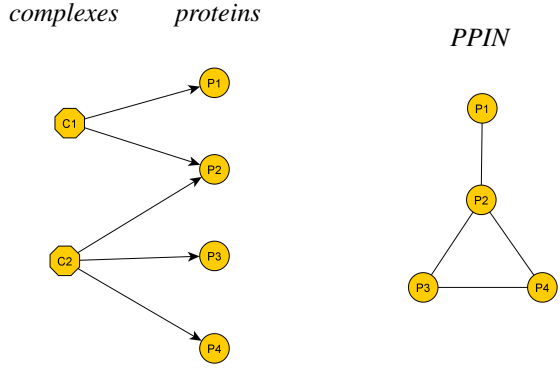

\begin{figure}[t]
\includegraphics[width=0.885\linewidth]{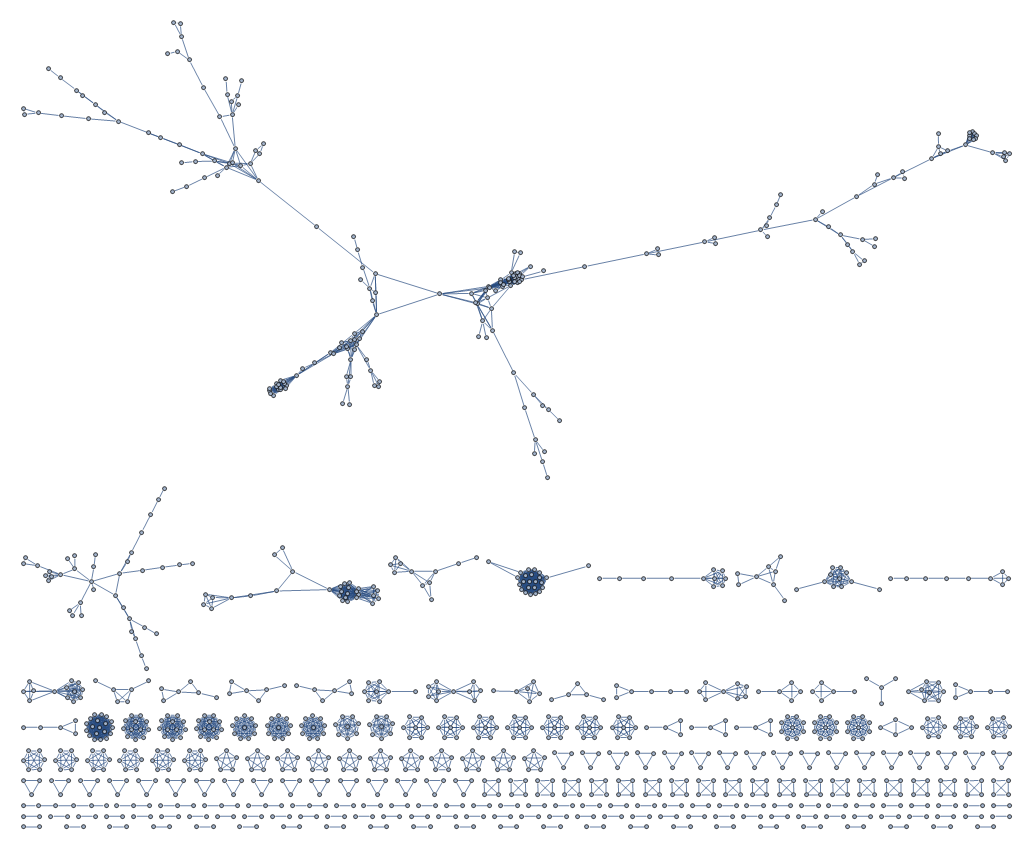}
\caption{Projection onto the protein space, in the sense of Figure \ref{fig:PPINcomplexes},  of the protein complexes and their core constituents identified in \emph{Sac. cerevisiae} \cite{gavin2006} via mass-spectroscopy. The resulting network shows distinct cliques, but only the beginning of the dense core that is generally seen in protein-protein interaction networks. One would expect the  dense core to emerge if the data included also non-core proteins, as well as protein reactions which may not necessarily correspond to named complexes. }
\label{fig:A-CGavin}
\end{figure} 

Many systems can be described with an interaction graph similar to the one of \cite{Barra, Barra_old}. For instance, although most protein-protein interaction network (PPIN) data repositories (such as \cite{hprd}) report only binary interactions between pairs of protein species,  the more natural description is in fact that of a bipartite graph, with one set of $\alpha N$ nodes representing protein complexes and another set representing the $N$ individual proteins. A link from a given protein to a given complex then indicates that the protein participates in the complex; see Figure \ref{fig:PPINcomplexes}.
If from such data we construct a new weighted graph, with nodes that represent the proteins only, and links between the nodes that give the number of complexes in which they jointly participate, we obtain once more graphs $\bc\in\graphset^\prime$ from  the graph ensemble (\ref{eq:separable_ensemble}). The only difference with the immune models is that now the variables $\xi_i^\mu$  take their values randomly and independently from the set $\{0,1\}$:
\begin{eqnarray}
p(\bc)=\Big\bra \prod_{i<j}\delta_{c_{ij},\sum_{\mu=1}^{p}\xi_i^\mu\xi_j^\mu}\Big\ket_{\bxi},~~~~~~~~
p(\xi_i^\mu) = \frac{q}{N} \delta_{\xi_i^\mu,1} + \left( 1 - \frac{q}{N} \right) \delta_{\xi_i^\mu,0} 
\label{eq:PPINseparable}
\end{eqnarray}
 This graph ensemble was studied in
 \cite{Newman2003}, in terms of its topological properties (percolation transition, path lengths, and so on), but not in the context of protein interactions. 
Several observables of the projected network $\bc\in\graphset^\prime$ can be immediately deduced from the parameters of the bipartite graph. For instance:
\begin{eqnarray}
\left\langle c_{ij} \right\rangle &=&\alpha N \left\langle \xi_i \xi_j \right\rangle = \alpha q^2/N
\\
 p(c_{ij}) &=& 
 \int_{-\pi}^\pi\! \frac{\rmd \omega}{2 \pi}
  \rme^{\rmi \omega c_{ij}}
   \left\langle \rme^{\rmi \omega \xi_i \xi_j} \right\rangle ^{\alpha N}
 _{ \left\lbrace \xi \right\rbrace} 
 ~=~\binom{\alpha N}{c_{ij}} \left[\frac{q^2}{N^2}\right]^{c_{ij}} \left[ 1 - \frac{q^2}{N^2}\right]^{\alpha N - c_{ij}}
\label{eq:probc}
 \end{eqnarray}
from which it also follows that
$\bar{k} = N^{-1} \sum_{ij} [ 1 - p(c_{ij}=0) ] = \alpha q^2 $.
As with the immune model,  one obtains loops in the PPIN graph $\bc$ again from an underlying model which is tree-like; here this underlying model  is the description of the system in terms of the proteins and their complexes. In Figure \ref{fig:A-CGavin} we show the result of using the data from \cite{gavin2006}  on protein complexes found in yeast in order to construct a protein-protein interaction network 
according to the definition in (\ref{eq:PPINseparable}), i.e. via $c_{ij}=\sum_{\mu}\xi_i^\mu\xi_j^\mu$, with $\xi_i^\mu$ indicating whether ($\xi_i^\mu=1$) or not ($\xi_i^\mu=0$) protein $i$ participates in complex $\mu$. It will be clear that, using the technology of \cite{Barra}, one should now be able to solve analytically models of protein reaction processes on `loopy' protein-protein interaction networks, 
provided they are built on these along the lines of (\ref{eq:PPINseparable}).

\section{Conclusion}

In this paper we have discussed some conceptual and mathematical issues that emerge as soon as one studies (processes on) graphs and networks that are not locally tree-like, but exhibit  extensive numbers of short loops. Stochastic processes on graphs with many short loops have quantitative characteristics that are bvery different from those that run on tree-like structures. This is why Ising models on tree-like lattices are trivially solved, but Ising models on finite-dimensional lattices are not. Yet our currently available arsenal of  mathematical tools for analysing `loopy' graphs is rather limited, because the tree-like assumption is at the very core of most of our techniques, whether we are analysing processes on graphs or calculating entropies of graph ensembles.  Even asymptotic results for infinitely large `loopy' graphs are largely absent. However, we believe that this area will become increasingly important in the coming years, since the fact is that many (if not  most) of the networks that we can observe in the real-world do have significant numbers of loops.  

We have tried to propose and explain some new ideas and possible  routes forward, aimed at making progress in the analytical study of `loopy' random graph ensembles.  Some built on our own ideas (including e.g. simple approximations and replica techniques) and some built on work of others (such as the papers by Burda et al, that are based on diagrammatic expansions). Most of the material relates to ongoing work, and is still only beginning to be explored. We discussed  in more detail two specific random graph ensembles: the Strauss model, because it is the simplest possible random graph ensemble in which one can induce arbitrary numbers of short loops (in this case triangles), and an ensemble that emerged in a recent immunological model, because this model could be solved analytically in spite of its `loopy' nature. 
\vsp

\noindent
{\bf Acknowledgements}
\\[2mm]
Both authors gratefully acknowledge financial support from the Biotechnology and Biological Sciences Research Council (BBSRC) of the United Kingdom, and would like to Franca Fraternali, Alessia Annibale and Eric Blanc for discussions and
Sun Sook Chung and Alessandro Pandini for the triangle count data used as the basis for Figure \ref{fig:strauss_vs_biology}.

%\bibliography{gd4}
%\bibliographystyle{plain}

\end{document}